\documentclass{emulateapj}
\usepackage{amsmath}
\usepackage{graphicx}
\usepackage [english]{babel}
\usepackage [autostyle, english = american]{csquotes}
\MakeOuterQuote{"}

\shorttitle{Luminous Extranuclear Star-forming Regions in LIRGs}
\shortauthors{S. Linden et al.}

\begin{document}

\title{A Very Large Array Survey of Luminous Extranuclear Star-forming Regions in Luminous Infrared Galaxies in GOALS}

\author{S. T. Linden\altaffilmark{1}, Y. Song\altaffilmark{1}, A. S. Evans\altaffilmark{1,2},  E. J. Murphy\altaffilmark{2}, L. Armus\altaffilmark{3}, L. Barcos-Mu\~noz\altaffilmark{2}, K. Larson\altaffilmark{3}, T. D\'iaz-Santos\altaffilmark{4}, G. C. Privon\altaffilmark{5}, J. Howell\altaffilmark{3}, J. A. Surace\altaffilmark{6}, V. Charmandaris\altaffilmark{7,8}, V. U\altaffilmark{9}, A. M. Medling\altaffilmark{10}, J. Chu\altaffilmark{11}, E. Momjian\altaffilmark{12}}


\altaffiltext{1}{Astronomy Department, University of Virginia, 530 McCormick Road, Charlottesville, VA 22904 USA}
\altaffiltext{2}{National Radio Astronomy Observatory, 520 Edgemont Road, Charlottesville, VA 22903 USA}
\altaffiltext{3}{Infrared Processing and Analysis Center, California Institute of Technology, MS 100-22, Pasadena, CA 91125 USA}
\altaffiltext{4}{Nucleo de Astronomia de la Facultad de Ingenieria, Universidad Diego Portales, Av. Ejercito Libertador 441, Santiago, Chile}
\altaffiltext{5}{Department of Astronomy, University of Florida, 211 Bryant Space Sciences Center, Gainesville, FL 32611}
\altaffiltext{6}{Eureka Scientific, Inc. 2452 Delmer Street Suite 100 Oakland, CA 94602-3017 USA}
\altaffiltext{7}{Department of Physics, University of Crete, GR-71003, Heraklion, Greece}
\altaffiltext{8}{Institute of Astrophysics, FORTH, GR-71110 Heraklion, Greece}
\altaffiltext{9}{Department of Physics and Astronomy, 4129 Frederick Reines Hall, University of California, Irvine, CA 92697, USA}
\altaffiltext{10}{Ritter Astrophysical Research Center, University of Toledo, Toledo, OH 43606, USA}
\altaffiltext{11}{Institute for Astronomy, University of Hawaii, Honolulu, HI 96822, USA}
\altaffiltext{12}{National Radio Astronomy Observatory, P.O. Box O, 1003 Lopezville Road, Socorro, NM 87801, USA}


\begin{abstract}
We present the first results of a high-resolution Karl G. Jansky Very Large Array (VLA) imaging survey of luminous and ultra-luminous infrared galaxies (U/LIRGs) in the Great Observatories All-Sky LIRG Survey (GOALS). From the full sample of 68 galaxies, we have selected 25 LIRGs that show resolved extended emission at sufficient sensitivity to image individual regions of star-formation activity beyond the nucleus.~With wideband radio continuum observations, which sample the frequency range from $3-33$ GHz, we have made extinction-free measurements of the luminosities and spectral indicies for a total of 48 individual star-forming regions identified as having de-projected galactocentric radii ($r_{G}$) that lie outside the 13.2$\mu$m core of the galaxy.~The median $3-33$ GHz spectral index and 33 GHz thermal fraction measured for these "extranuclear" regions is $-0.51 \pm 0.13$ and $65 \pm 11\%$ respectively.~These values are consistent with measurements made on matched spatial scales in normal star-forming galaxies, and suggests that these regions are more heavily-dominated by thermal free-free emission relative to the centers of local ULIRGs.~Further, we find that the median star-formation rate derived for these regions is $\sim 1 M_{\odot}$ yr$^{-1}$, and when we place them on the sub-galactic star-forming main sequence of galaxies (SFMS), we find they are offset from their host galaxies' globally-averaged specific star-formation rates (sSFRs).~We conclude that while nuclear starburst activity drives LIRGs above the SFMS, extranuclear star-formation still proceeds in a more extreme fashion relative to what is seen in local spiral galaxies.


\end{abstract}

\keywords{galaxies: interactions - infrared: galaxies - H II regions - radio continuum: general - stars: formation}

\section{Introduction}

Galaxies with high infrared (IR) luminosities, e.g., luminous infrared galaxies (LIRGs: $10^{11} < L_{\rm IR} [8 - 1000 \mu{\rm m}] < 10^{12}$ L$_\odot$), are rare in the local Universe, yet they are a cosmologically important class of objects that dominate the infrared luminosity density at redshifts $z=1-2$ (e.g. Murphy et al. 2011; Magnelli et al. 2013). LIRGs, which are often triggered by the interactions and mergers of gas-rich disk galaxies, have high bolometric luminosities that primarily emanate from nuclear star-formation, as well as active galactic nuclei (AGN) (e.g., Sanders \& Mirabel, 1996). In addition to strong nuclear star formation, enhanced star-formation activity has been seen in the outer disks and tidal structures of many interacting galaxies (Schweizer 1978; Hibbard \& van Gorkom 1996; Smith et al. 2010; Smith et al. 2016). 


To fully assess the nature of star formation and AGN activity in LIRGs as a function of merger stage, luminosity, and optical depth, we initiated the Great Observatories All-sky LIRG Survey (GOALS; Armus et al. 2009). The multi-wavelength dataset, which is most complete for LIRGs with $L_{IR} \geq 10^{11.4} L_{\odot}$, contains observations using Chandra (Iwasawa 2011; Torres-Alba 2018), GALEX (Howell et al. 2010), HST (Kim et al. 2013; Linden et al. 2017), Spitzer (Stierwalt et al. 2013, 2014), Herschel (e.g., D\'iaz-Santos et al. 2013, 2017; Chu et al. 2017), VLA (e.g. Barcos-Munoz et al. 2015, 2017) and ALMA (e.g., Xu et al. 2014, 2015; Privon et al. 2017). 

These studies have revealed one of the primary challenges in studying LIRGs is that their nuclear regions can be heavily dust-enshrouded, thus necessitating long wavelength observations to unveil their properties (e.g., Condon et al. 1991; Downes \& Solomon 1998; Soifer et al. 2000). Although weak with respect to a galaxies' bolometric luminosity, radio emission is largely optically-thin and unaffected by dust extinction. The emission is primarily powered by stars more massive than $\sim 8 M_{\odot}$ which end their lives as core-collapse supernovae, and their remnants are thought to be the primary producers of cosmic ray (CR) electrons that give rise to the diffuse synchrotron emission observed from star-forming galaxies (Condon et al. 1992). These same massive stars are also responsible for the creation of H II regions that produce radio free-free emission, whose strength is directly proportional to the production rate of ionizing (Lyman continuum) photons. 


Ground-observable radio frequencies ($\sim 1-100$ GHz) are particularly useful in probing both processes. The nonthermal component typically has a steep spectrum [$S(\nu) \propto \nu^{\alpha_{NT}}$, where $\alpha_{NT} \sim -0.85$], while the thermal (free-free) component is relatively flat ($\alpha_{T} \sim -0.1$; e.g., Condon et al. 1992). Accordingly, for globally integrated measurements of star-forming galaxies, lower frequencies (e.g., 1.4 GHz) are generally dominated by nonthermal emission, while the observed fraction of thermal free-free emission increases with frequency, eventually dominating beyond $\sim$ 30 GHz. Thus, by observing galaxies across this frequency range we can separate these two emission components and produce maps of the current star formation activity.

However, a robust decomposition of the radio spectral energy distribution (SED) may be complex. For example: the thermal and nonthermal fractions may vary with galaxy mass (e.g., Hughes et al. 2007, Bell 2003), the nonthermal spectral index can vary within galaxies (Tabatabaei et al. 2013, 2017), and anomalous microwave emission may add unexpected features to the radio SED in some regions (e.g., Murphy et al. 2011, 2018b). More relevant to the present study, observations of U/LIRGs over the last decade have revealed that they can have significant variation from system-to-system in their global radio properties (Leroy et al. 2011; Murphy et al. 2013; Barcos-Munoz et al. 2017). 

This was the basis for the GOALS "equatorial" survey, which is a multi-frequency VLA program to image a complete sample of 68 U/LIRGs within the declination range $-20^{o} < \delta < 20^{o}$, from $3-33$ GHz at resolutions of $10-1000$ pc. One of the fundamental goals of this study is to quantify the level of variation we see in the radio SEDs on sub-galactic scales in these galaxies, and determine the validity of applying a two-component power-law model to characterize the star-formation activity of individual regions in LIRGs.

In this paper we examine the radio and near-IR properties of "extranuclear" star-forming regions identified in galaxies in the GOALS equatorial sample.~This spatial cut is imposed to control for any contribution to the observed radio continuum emission from a strong central AGN. Thus, we require all regions for which we perform photometric analysis to reside outside the measured mid-infrared (MIR) core of the galaxy, where an AGN would have the largest influence (see Section 3.2 for details). Further, due to the sensitivity of our VLA observations, we are only able to observe the most luminous star-forming regions in the disks of these galaxies. An analysis of the radio continuum properties of nuclear and circumnuclear star-forming regions in LIRGs will be presented in a series of future papers.


The paper is organized as follows:~In Section 2 we describe the observations, data reduction, imaging, and sub-sample selection. In Section 3, we describe our ancillary Spitzer data products, and outline our method for identifying individual extranuclear star-forming regions and extracting multi-wavelength photometry. In Section 4, we discuss the radio continuum properties of the sample. In Section 5, we discuss these results in the context of both the far-infrared - radio correlation, as well as the star-formation rate main sequence. In Section 6 we summarize the results.

Throughout this paper, we adopt WMAP Three-Year Cosmology of $H_{0} = 73$ km s$^{-1}$ Mpc$^{-1}$, $\Omega _{\rm matter} = 0.27$, and $\Omega _{\Lambda} = 0.73$ (Spergel et al. 2007).

\section{Sample and Data Analysis}

\begin{deluxetable*}{lrrccccc}
\tabletypesize{\footnotesize}
\tablewidth{0pt}
\tablecaption {Properties of the 25 GOALS Galaxies in the sub-sample}
\tablehead{
\colhead{Name} & \colhead{RA} & \colhead{Dec} & \colhead{D(Mpc)} & \colhead{Log($L_{IR}/L_{\odot}$)} & \colhead{Core FWHM (kpc)\tablenotemark{a}} & \colhead{Inc ($^{\circ}$)\tablenotemark{b}} & \colhead{PA ($^{\circ}$)\tablenotemark{b}}} \\
\startdata
MCG-02-01-052 & 00h18m50.10s & -10d22m42.0s & 105.76 & 11.41 & 2.28 & 0.35 & 55 \\
IC1623 & 01h07m47.49s & -17d30m25.3s & 78.57 & 11.65 & 4.60 & 0.39 & 123 \\
MCG-03-04-014 & 01h10m08.96s & -16d51m09.8s & 136.17 & 11.63 & 4.16 & 0.85 & 64 \\
NGC0838 & 02h09m38.56s & -10d08m49.1s & 50.13 & 11.00 & 1.65 & 0.15 & 84 \\
IC0214 & 02h14m05.59s & +05d10m23.7s & 117.29 & 11.37 & 4.79 & 0.76 & 139 \\
NGC0877 & 02h17m59.64s & +14d32m38.6s & 50.29 & 11.04 & 5.82 & 0.11 & 175 \\
UGC02238 & 02h46m17.49s & +13d05m44.4s & 83.37 & 11.26 & 3.64 & 0.52 & 135 \\
UGC02369 & 02h54m01.78s & +14d58m14.0s & 121.94 & 11.60 & 3.52 & 0.75 & 30 \\
CGCG465-012 & 03h54m16.08s & +15d55m43.4s & 86.84 & 11.15 & 2.61 & 0.14 & 89 \\
UGC02982 & 04h12m22.45s & +05d32m50.6s & 67.57 & 11.13 & 3.11 & 0.46 & 106 \\
UGC03094 & 04h35m33.83s & +19d10m18.2s & 97.06 & 11.35 & 4.53 & 0.39 & 179 \\
IRAS05442+1732 & 05h47m11.18s & +17d33m46.7s & 74.95 & 11.25 & 1.56 & 0.52 & 70 \\
IC0563 & 09h46m20.30s & +03d02m44.0s & 87.01 & 11.19 & 2.84 & 0.38 & 107 \\
NGC3110 & 10h04m02.11s & -06d28m29.2s & 73.48 & 11.31 & 2.96 & 0.32 & 176 \\
IC2810 & 11h25m45.05s & +14d40m35.9s & 142.89 & 11.59 & 3.72 & 0.39 & 32 \\
NGC5257 & 13h39m52.90s & +00d50m24.0s & 98.63 & 11.55 & 8.98 & 0.70 & 96 \\
NGC5258 & 13h39m57.70s & +00d49m51.0s & 98.63 & 11.55 & 9.60 & 0.36 & 169 \\
NGC5331 & 13h52m16.20s & +02d06m03.0s & 139.1 & 11.59 & 5.25 & 0.35 & 102 \\
NGC5936 & 15h30m00.84s & +12d59m21.5s & 60.81 & 11.07 & 1.33 & 0.76 & 60 \\
NGC5990 & 15h46m16.37s & +02d24m55.7s & 58.42 & 11.06 & 1.22 & 0.60 & 99 \\
CGCG052-037 & 16h30m56.54s & +04d04m58.4s & 104.72 & 11.38 & 2.49 & 0.14 & 114 \\
IRASF16516-0948 & 16h54m24.03s & -09d53m20.9s & 96.87 & 11.24 & 5.40 & 0.18 & 110 \\
IRASF17138-1017 & 17h16m35.60s & -10d20m38.0s & 75.84 & 11.42 & 3.64 & 0.59 & 83 \\
NGC7592 & 23h18m22.54s & -04d24m58.5s & 95.13 & 11.33 & 3.76 & 0.40 & 90 \\
NGC7679 & 23h28m46.66s & +03d30m41.1s & 67.7 & 11.05 & 1.89 & 0.19 & 94
\enddata
\tablenotetext{a}{13.2$\mu$m core sizes taken from D\'iaz-Santos et al. (2010).}
\tablenotetext{b}{Inclinations and Position Angles taken from Kim et al. (2013), Paturel et al. (2003), and Jarrett et al. (2000)}
\end{deluxetable*}

\subsection{VLA Observations and Data Reduction}
The VLA observations were carried out for the full sample during three separate C-configuration cycles in February 2014 (14A-471), February 2016, and March 2016 (16A-204). The maximum baseline of this array configuration is 3.4 km. Three VLA receiver bands were utilized in these observations: the S-band (2-4 GHz), Ku-band (12-18 GHz), and Ka-band (26.5-40 GHz). The 8-bit samplers were used for the S-band observations, delivering 2 GHz of total bandwidth, centered at 3 GHz, by using two 1 GHz baseband pairs, both with right- and left hand circular polarizations. The 3-bit samplers were used in the Ku-band observations, delivering 6 GHz of total bandwidth, centered at 15 GHz, by using three 2 GHz baseband pairs. The 3-bit samplers were also used in the Ka-band observations, delivering 8 GHz of total bandwidth, centered at 33 GHz, by using four 2 GHz baseband pairs.~Hereafter, we may refer to the data and/or the images from the S-, Ku-, and Ka-bands as the 3, 15, and 33 GHz data and/or images, respectively.

To reduce and calibrate the VLA data we followed standard calibration and editing procedures, and utilized the VLA calibration pipeline built on the Common Astronomy Software Applications (CASA; McMullin et al. 2007) versions 4.6.0 and 4.7.0. After each pipeline run, we manually inspected the visibilities and calibration tables for evidence of bad antennas, frequency ranges, or time ranges, flagging correspondingly. We also flagged any instances of RFI, for which we removed several significantly affected frequency ranges in the S- and Ku-bands, and very little of the 33 GHz data. After flagging, we re-ran the pipeline, and repeated this process until we could not detect any further signs of poorly-calibrated data. A detailed description of our data reduction procedures can be found in Murphy et al. (2018a).

\subsection{Galaxy Selection}
The goal of this study is to resolve individual extranuclear regions within LIRGs for which we can perform multi-band photometry, and extract detailed information about their star-formation properties.

\begin{figure}
\centering
\includegraphics[scale=0.43]{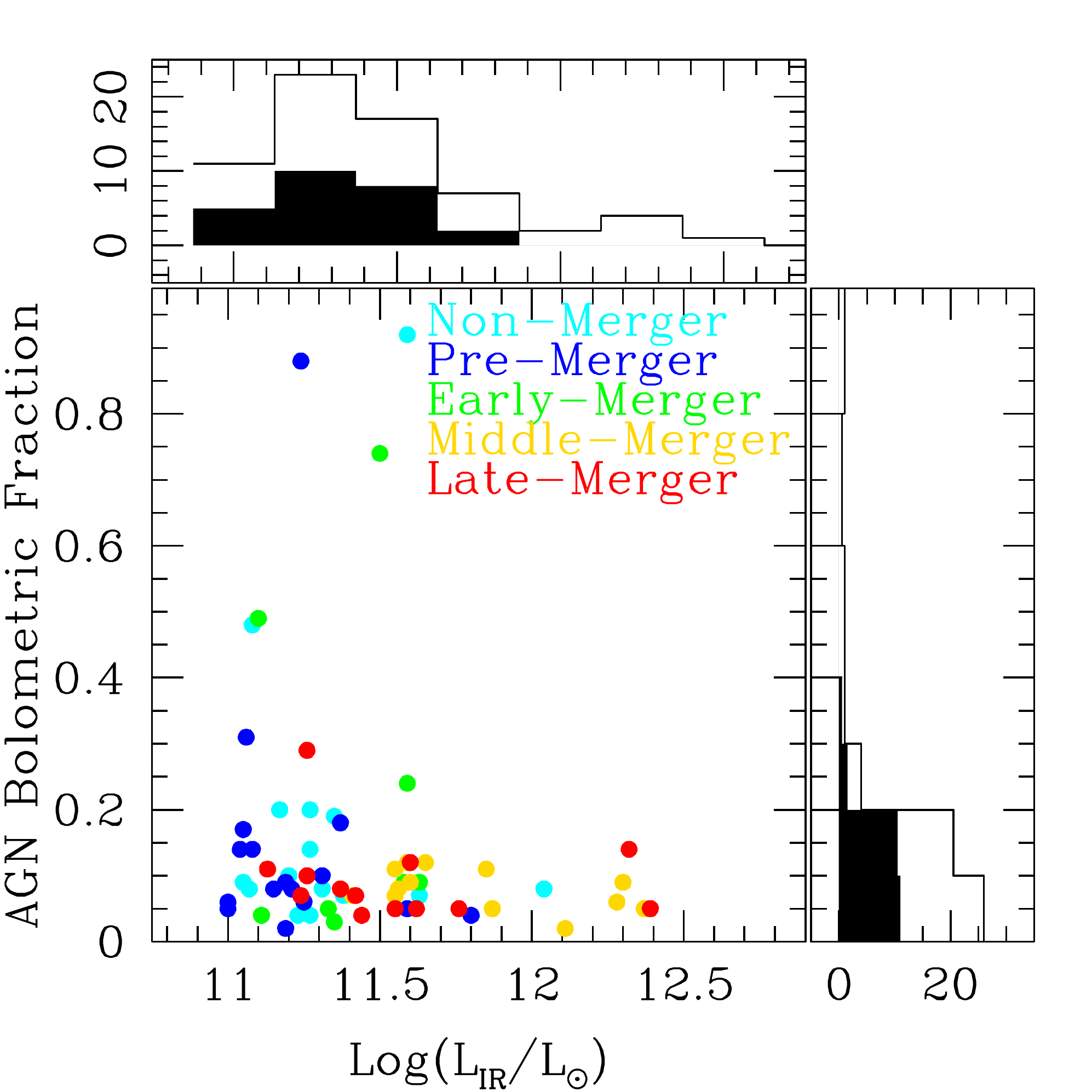}
\caption{The distribution of infrared luminosity and bolometric AGN fraction all galaxies in GOALS within $+20$ and $-20$ degrees declination. The AGN fractions are taken from a recent compilation in Diaz-Santos et al. (2017), and the galaxies are color-coded by the observed merger stages taken from Haan et al. (2013) and Stierwalt et al. (2013). The histograms on each side show the full distribution of equatorial GOALS sources, with the black shaded histogram indicating the sub-sample selected for this study. What is clear is nearly all of the galaxies identified in our sub-sample are classified starburst-dominated systems.}
\end{figure}

We therefore selected from the full sample of 68 equatorial LIRGs the 25 which showed resolved extended structure with the VLA across all three frequency bands (Table 1). By targeting galaxies with high $L_{IR}$, the GOALS sample includes the most extreme starbursts and AGN in the local Universe. Importantly, these galaxies are different from the normal star-forming galaxies studied previously on resolved scales in the local Universe (Alonso-Herrero \& Knapen 2001; Bradley et al. 2006; Liu et al. 2013; Smith et al. 2016). However, in order to properly place our results in the greater context of galaxy evolution, we must also be able to disentangle any contribution of a strong central AGN to the measured luminosities of individual regions.

From our sub-sample of 25 galaxies we see that the majority (22/25) of the extended galaxies in our VLA survey are indeed starburst-dominated, with AGN bolometric fractions under 15$\%$ (Diaz-Santos et al. 2017). Figure 1 shows that our sub-sample of sources span both the complete range of interaction stages (i.e widely separated disk galaxies to fully merged systems: Haan et al. 2013; Stierwalt et al. 2013), and a luminosity range of $L_{IR} = 10^{11.00-11.65} L_{\odot}$. The final analysis of extranuclear star-forming regions within this sub-sample allows for direct comparison of the luminosity, SFR, radio spectral indices, and overall morphologies of the active star-forming regions in LIRGs and normal spiral galaxies without issues associated with contamination from an AGN or obscuration due to dust.

\subsection{Interferometric Imaging}

Calibrated VLA measurement sets for each source were imaged using the task tclean in CASA version 4.7.0. The mode of tclean was set to multi-frequency synthesis (mfs; Conway et al. 1990; Sault \& Wieringa 1994). For nearly all sources, we chose to use Briggs weighting with robust=0.5, and set the variable nterms=2, which allows the cleaning procedure to additionally model the spectral index variations on the sky. To help deconvolve extended low-intensity emission, we took advantage of the multiscale clean option (Cornwell 2008; Rau \& Cornwell 2011) in CASA, searching for structures with scales $\sim1$ and 3 times the full width at half-maximum (FWHM) of the synthesized beam. For our S- and Ku-band data we also implemented the widefield uv-plane gridding and the w-projection algorithm with 16 planes to better model the curvature of the low-frequency sky. For those sources where Briggs weighting of robust=0.5 failed to capture any significant emission or structure, we increased this factor in steps of 0.5 towards natural weighting (i.e. robust=2.0) until the source was detected, and the sidelobes were sufficiently suppressed.~A summary of the imaging parameters is given in Table 2.


\begin{deluxetable}{llll}
\tabletypesize{\footnotesize}
\tablecolumns{4}
\tablewidth{0pt}
\tablecaption {Imaging Parameters}
\tablehead{
\colhead{Inputs} & \colhead{Ka-Band} & \colhead{Ku-Band} & \colhead{S-Band}} \\
\startdata
Frequency & 33 GHz & 15 GHz & 3 GHz \\
Cell\tablenotemark{a} & 0.12 & 0.27 & 0.12 \\
UV-Gridder & Widefield & Widefield & Widefield \\
Multiscales\tablenotemark{b} & 0,10,30 & 0,10,30 & 0,10,30 \\
Nterms\tablenotemark{c} & 2 & 2 & 2  \\
Robust\tablenotemark{d} & 1.0 or 0.5 & 1.0 or 0.5 & 0.5 \\
\enddata
\tablenotetext{a}{The cell size is given in arcseconds/pixel}
\tablenotetext{b}{The scale size is given in pixels, with 0 being a point source.}
\tablenotetext{c}{Number of terms used in Taylor series expansion}
\tablenotetext{d}{The robust weighting scheme was chosen for all images}
\end{deluxetable}

A primary beam correction was further applied using the CASA task impbcor before analyzing the images. The primary-beam-corrected continuum images at each frequency for two of the targets used in this study are shown in Figure 2. Finally, in order to measure flux densities and spectral indices within consistent apertures across all three VLA bands, we convolve the images to a common circularized beam that is closest in size to the image with the lowest resolution for each galaxy. This allows us to make the cleanest comparison of structure across multiple bands, and minimize the effect of using circular apertures to extract photometry. The FWHM of the smoothed beam size for each galaxy is given in Table 3, along with the corresponding point-source rms sensitivity of each image.

\begin{figure*}
\centering
\includegraphics[scale=0.6]{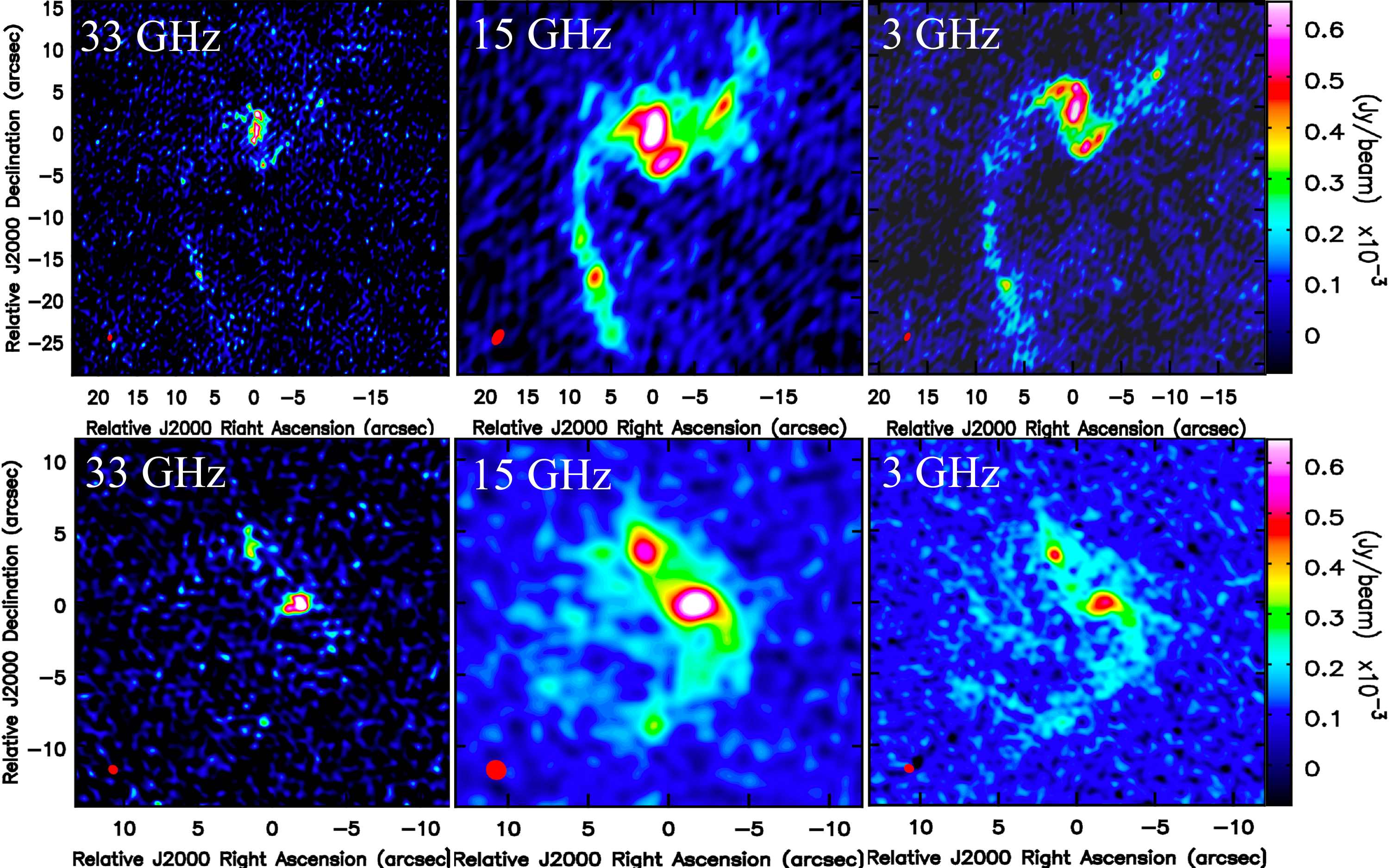}
\caption{Our native-resolution multi-band VLA imaging at 33, 15, and 3 GHz from left to right. The top three panels show CGCG 465-012, and the bottom three panels show NGC 3110. The colormap scaling for each image is given on the right. The solid red ellipse in the bottom corner is the beam size of each image, with angular resolutions of $\sim$ 0.7'', 1.4'', and 0.7'' respectively. These galaxies were chosen as representative cases for the sample, as they live at the median distance of the sub-sample ($\sim$86 Mpc).}
\end{figure*}

\section{Ancillary Data and Region Photometry}

\subsection{Spitzer IRAC Imaging}

With the addition of near-IR Spitzer imaging at 3.6, 4.5, 5.8, and $8\mu$m, we can make direct comparisons of the measured 33 GHz SFR, powered by young massive stars, to the total stellar mass ($M_{*}$) inferred from the evolved low-mass stellar populations within the same regions. Additionally, we can estimate the total infrared luminosity ($L_{IR}$ [8-1000$\mu$m]) per region, and compare these results to the observed low-frequency ($\sim 3$GHz) radio emission, which has been shown to serve as a reliable proxy of IR emission in local star-forming galaxies via the well-studied IR-radio correlation (Helou 1985; Condon et al. 1992).


Spitzer IRAC channels (Ch) $1-4$ data were taken as part of GOALS, and details on the associated observation strategies and data reduction steps are available in Mazzarella et al. (2019 in prep). IRAC channels $3-4$ (5.8 and 8.0 $\mu$m) are primarily sensitive to emission due to polycyclic aromatic hydrocarbons (PAHs; e.g., Leger \&, Puget 1984), whereas Spitzer/IRAC NIR channels 1 and 2 (3.6 and 4.5$\mu$m) data are treated as free of hot dust emission, except when a powerful AGN is present, and thus primarily sensitive to old stellar emission (e.g., Helou et al. 2004). Lu et al. (2003) used global measurements of nearby star-forming galaxies with ISO, which spanned 3 orders of magnitude in IR-luminosity, to confirm that hot dust does not contribute significantly to the emission below 3 $\mu$m. Hunt et al. (2002) further showed that the contribution in Spitzer Ch $1-2$ is on average $3-4\%$, making these bands highly sensitive probes of stellar emission. We therefore utilize the calibration presented in Querejeta et al.~(2015) to convert our 3.6 and 4.5$\mu$m flux densities to total stellar mass ($M_{*}$).

To account for the significant fraction of scattered light in the images due to the structure of the Spitzer point-spread function (PSF), we use the convolution kernels presented in Aniano et al.~(2011) to deconvolve the Spitzer PSF from each image, and produce corrected images for each band. 


\begin{deluxetable*}{lcrccc}
\tabletypesize{\footnotesize}
\tablecolumns{6}
\tablewidth{0pt}
\tablecaption {Source Imaging Characteristics}
\tablehead{
\colhead{Name} & \colhead{Program ID} & \colhead{$\theta_{s}$ (arcsec)\tablenotemark{a}} & \colhead{$\sigma_{33}$ (mJy/beam)\tablenotemark{b}} & \colhead{$\sigma_{15}$ (mJy/beam)\tablenotemark{b}} & \colhead{$\sigma_{3}$ (mJy/beam)\tablenotemark{b}}} \\
\startdata
MCG-02-01-051 & 14A-471 & 2.071 x 2.071 & 3.253 & 2.193 & 0.622 \\
IC1623 & 14A-471 & 3.018 x 3.018 & 0.337 & 0.135 & 0.833 \\
MCG-03-04-014 & 14A-471 & 2.917 x 2.917 & 0.180 & 0.064 & 0.268 \\
NGC0838 & 14A-471 & 2.846 x 2.846 & 0.282 & 0.096 & 0.439 \\
IC0214 & 14A-471 & 2.171 x 2.171 & 0.073 & 0.039 & 0.139 \\
NGC0877 & 14A-471 & 1.851 x 1.851 & 0.064 & 0.035 & 0.060 \\
UGC02238 & 16A-204/14A-471 & 1.867 x 1.867 & 0.061 & 0.038 & 0.097 \\
UGC02369 & 16A-204/14A-471 & 1.806 x 1.806 & 0.066 & 0.046 & 0.182 \\
CGCG465-012 & 16A-204/14A-471 & 1.760 x 1.760 & 0.042 & 0.021 & 0.049 \\
UGC02982 & 16A-204/14A-471 & 1.878 x 1.878 & 0.051 & 0.025 & 0.090 \\
UGC03094 & 14A-471 & 1.746 x 1.746 & 0.053 & 0.027 & 0.044 \\
IRAS05442+173 & 14A-471 & 1.759 x 1.759 & 0.091 & 0.065 & 0.129 \\
IC0563 & 14A-471 & 2.086 x 2.086 & 0.071 & 0.025 & 0.055 \\
NGC3110 & 14A-471 & 2.431 x 2.431 & 0.091 & 0.045 & 0.146 \\
IC2810 & 14A-471 & 2.169 x 2.169 & 0.065 & 0.040 & 0.074 \\
NCG5257 & 16A-204/14A-471 & 1.953 x 1.953 & 0.040 & 0.025 & 0.057 \\
NGC5258 & 16A-204/14A-471 & 1.953 x 1.953 & 0.028 & 0.075 & 0.101 \\
NCG5331 & 16A-204/14A-471 & 1.914 x 1.914 & 0.045 & 0.030 & 0.191 \\
NGC5936 & 16A-204/14A-471 & 1.749 x 1.749 & 0.069 & 0.039 & 0.086 \\
NGC5990 & 16A-204/14A-471 & 1.860 x 1.860 & 0.070 & 0.047 & 0.060 \\
CGCG052-037 & 16A-204/14A-471 & 1.854 x 1.854 & 1.147 & 0.031 & 0.088 \\
IRASF16516-09 & 16A-204/14A-471 & 2.334 x 2.334 & 0.091 & 0.040 & 0.262 \\
IRASF17138-10 & 16A-204/14A-471 & 2.292 x 2.292 & 0.127 & 0.067 & 0.210 \\
NGC7592 & 14A-471 & 2.116 x 2.116 & 0.141 & 0.078 & 0.075 \\
NGC7679 & 14A-471 & 1.849 x 1.849 & 0.060 & 0.030 & 0.100
\enddata
\tablenotetext{a}{The highest achieved resolution across our three VLA bands}
\tablenotetext{b}{These values represent the PSF rms sensitivities of each image}
\end{deluxetable*}

\subsection{Extranuclear Region Identification}

Extranuclear candidate regions were identified as being discrete knots with a $S/N$ of at least 3 in all three radio bands before a background subtraction of the surrounding diffuse radio emission within the galaxy. The median $S/N$ of the regions selected for photometry is $\sim$10, 37, and 15 for 33 GHz, 15 GHz, and 3 GHz respectively. A full list of regions and their photometric properties is given in the Appendix. To determine the physical separation for each of the identified candidate regions we compiled the axis-ratio ($b/a$) and position angle (PA) of all of the host galaxies (see Table 1). We adopted a hierarchy whereby values are taken from our HST-GALFIT structural analysis (Kim et al. 2013) were used first, then values were taken from the latest release of the HYPERLEDA database (Paturel et al. 2003), and finally if necessary we take values from the 2MASS Extended Source Catalog (Jarrett et al. 2000). We then used the equation in Dale et al. (1997) to calculate inclination angle (i) of each galaxy such that,

\begin{equation} 
\cos^{2}(i) = \frac{(b/a)^{2} - (b/a)^{2}_{int}}{1 - (b/a)^{2}_{int}}
\end{equation}

\noindent where $a$ and $b$ are the observed semi-major and semi-minor axes. The intrinsic axial ratio ($(b/a)_{int}$) is 0.2 for morphological types earlier than Sbc, and 0.13 otherwise. These measurements allow us to convert the apparent angular separation of a region to its host galaxy nucleus into a de-projected galactocentric radius ($r_{G}$) in units of kpc. We then compared all the measured $r_{G}$ values of the candidate extranuclear regions to the $13.2\mu$m core sizes (FWHM), as measured from D\'iaz-Santos et al. (2010) using Spitzer/IRS 2D spectra, which serve as our best probe of the total size of the nuclear regions in these systems. Three galaxies in our sub-sample (IC 1623, NGC 7592, NGC 7679) were not studied in that paper, and therefore we estimate the core-size using the IRAC $8\mu$m imaging as a proxy. We note that $8\mu$m emission is more extended than the MIR continuum in a significant fraction of LIRGs, making our estimates conservative upper-limits (D\'iaz-Santos et al. 2011). Of the 50 candidates we manually identified 48 regions in 25 galaxies with $r_{G}$ values larger than the MIR core radius, and were thus retained in the final sample of extranuclear star-forming regions.

\subsection{Aperture Photometry}

In order to measure beam-matched photometry between the VLA and Spitzer images, we smoothed all VLA, 3.6, and 4.5$\mu$m data to a common gaussian beam with a FWHM of $2.5"$ (the best resolution achievable across all 5 bands). This resulted in a physical resolution of $\sim 1$ kpc for the median distance to the galaxies in the sample (86 Mpc). In Figure 3, the beam-matched images for all three VLA bands as well as the IRAC Ch1 data are shown for the same sources as in Figure 2. 

We chose an aperture diameter of $4"$, which is larger than the FWHM of the worst resolution for any galaxy in the sample. To extract consistent photometry for the 8$\mu$m data, we applied an empirical aperture correction from Reach et al.~(2005) to account for missing flux due to the irregular shape of the IRAC PSF. We verified these empirical corrections gave consistent results when compared with the photometry done on the de-convolved 3.6$\mu$m and 4.5$\mu$m data alone. For the VLA imaging, the RMS was calculated by scaling the measured uncertainties by the square root of the restoring-beam area to the aperture area to account for the size of the aperture used. Finally, we take the intrinsic VLA flux uncertainty of $\sim 3\%$ added in quadrature with our empirically measured noise (Perley \& Butler 2013).

In order to account for the fact that many of these regions are not isolated, but rather, are embedded in larger star-forming disks, we perform a local background subtraction (using an annulus of $1"$ surrounding the photometric aperture) for each source in both the VLA and IRAC imaging. This allows us to separate the radio and IR emission directly associated with the young star-forming regions we identify, and is the only way to consistently compare the two wavelength regimes together. The resulting background-subtracted measurements and uncertainties for each region are given in the Appendix Table A.2. All further derived quantities and results are based on our background-subtracted photometry.



\begin{figure*}
\centering
\includegraphics[scale=0.6]{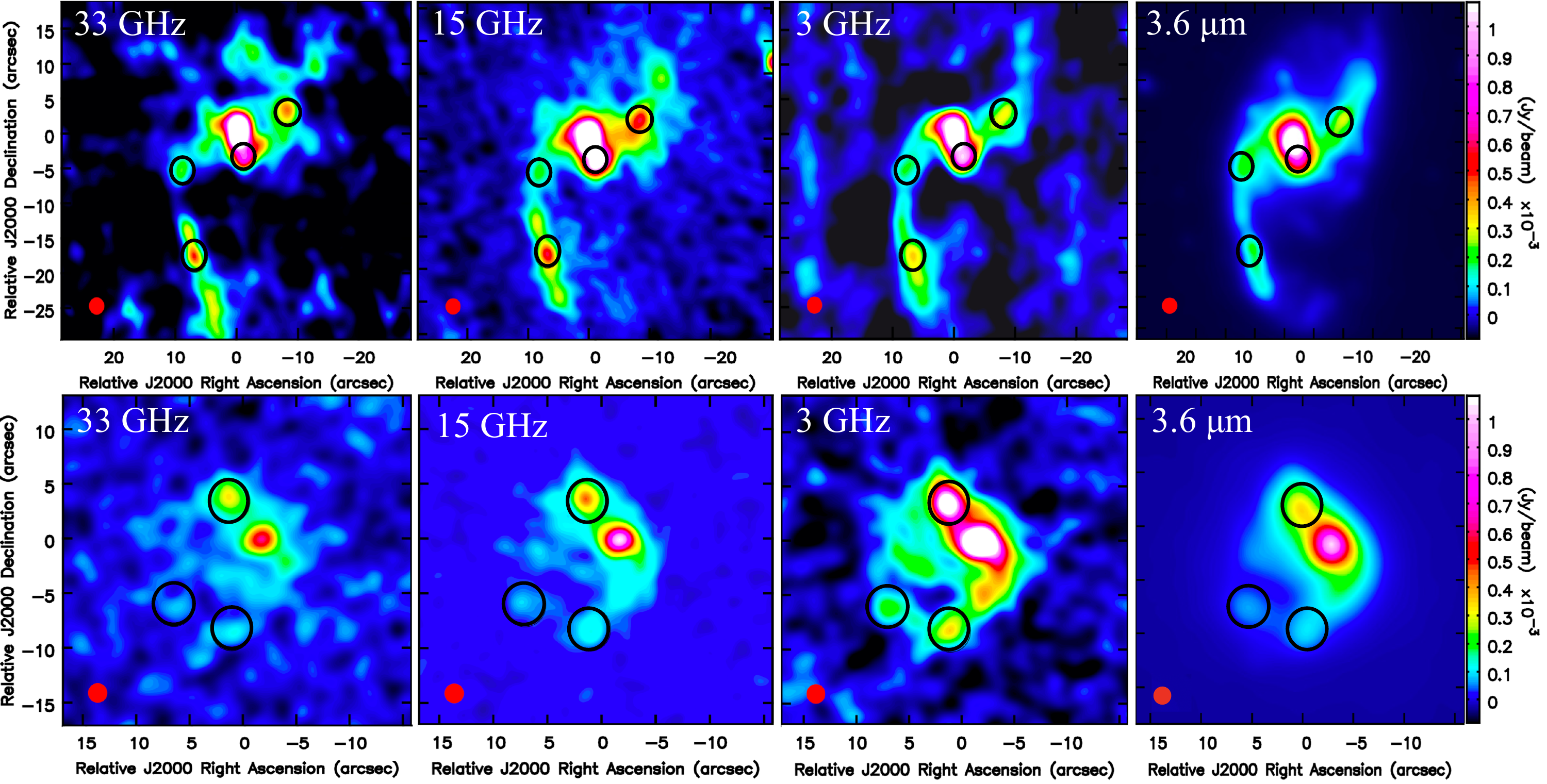}
\caption{Our matched-resolution multi-band VLA imaging at 33, 15, 3 GHz, and 3.6$\mu$m IRAC Ch1 from left to right. The top four panels show CGCG 465-012, and the bottom four panels show NGC 3110. The colormap scaling for each image is given on the right. The circularized red beam in the bottom left corner is matched across all 4 images, and is $\sim$ 1.7'' and 2.4'' for the top and bottom panels respectively. The black circles represent our photometric apertures centered on our confirmed extranuclear star-forming regions. These galaxies were chosen as representative cases for the sample, as they live at the $\sim$ median distance of the sub-sample (86 Mpc).}
\end{figure*}

\section{Results}

With matched-resolution observations at $\sim 2.5"$, which cover three broad windows within the 3 - 33 GHz frequency range, we can measure the spectral slope, the relative contributions of thermal and nonthermal emission, the star-formation rates, and ages of individual star-forming regions within our subsample of equatorial LIRGs.

\subsection{Radio Spectral Indices}

When interpreting the observed radio SEDs of galaxies we adopt a two-component power-law, with the thermal/nonthermal ratio as well as the nonthermal spectral index set as free parameters. For many normal and extreme star-forming galaxies in the local Universe this model adequately describes the dominant physical processes occurring (Condon et al. 1992; Murphy et al. 2012a). However for U/LIRGs, these models have mainly been applied to globally-integrated measurements, and scarcely studied on sub-galactic scales (Scoville et al. 2017).

\begin{figure}
\centering
\includegraphics[scale=0.43]{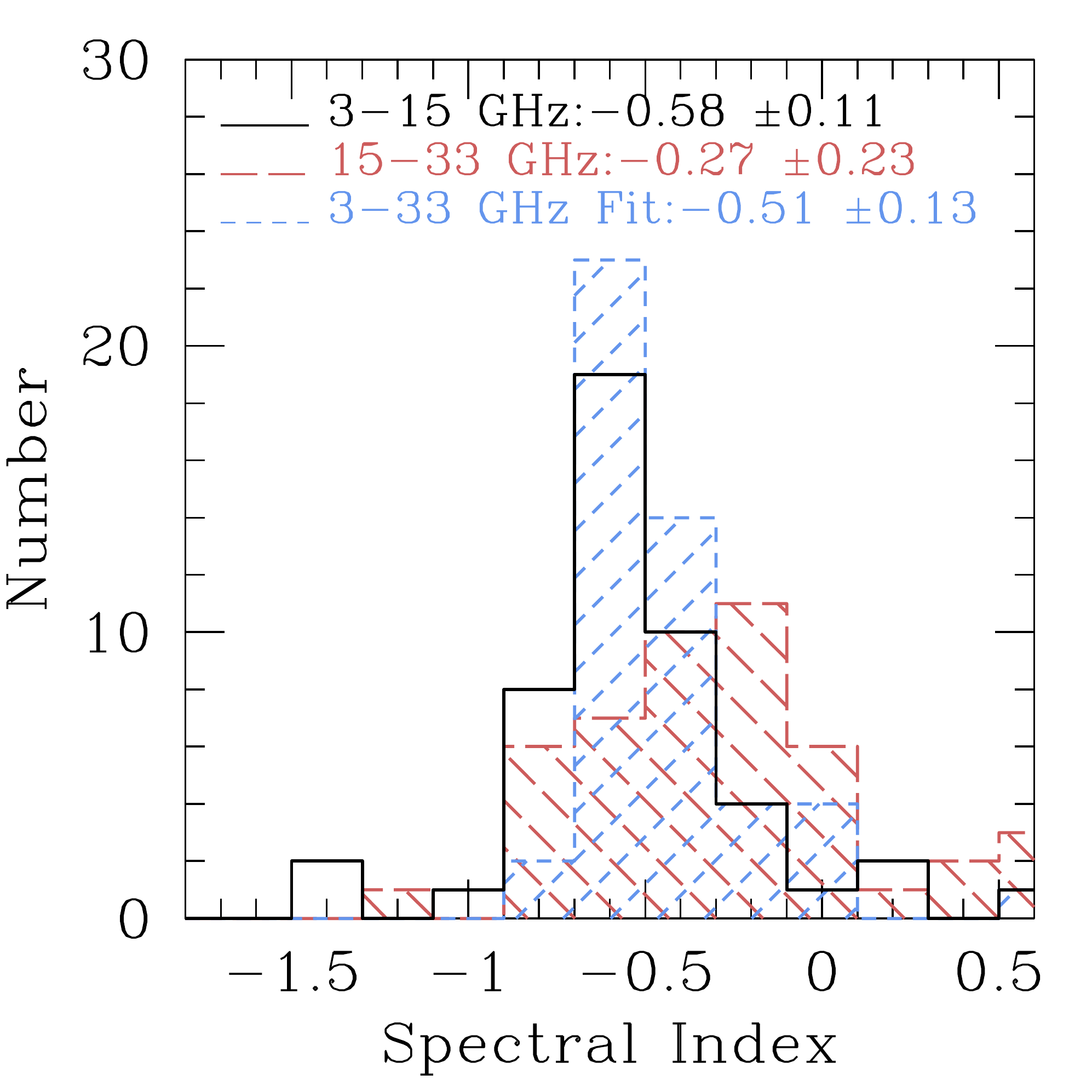}
\caption{The measured spectral index values for all extranuclear regions identified in the GOALS sample. Histograms of the inter- ($3 - 15$ and $15 - 33$ GHz) and full-band ($3 - 33$ GHz) spectral index distributions are shown in blue short-dashed, solid black, and red long-dashed lines respectively. The median values of $-0.58 \pm 0.11$ and $-0.27 \pm 0.23$ for the $3 - 15$ and $15 - 33$ GHz spectral index distributions indicates an increasing contribution of thermal free-free emission to the radio SED at increasing frequencies.}
\end{figure}

To measure the $3 - 33$ GHz spectral indices we performed a linear least-squares fit to the data with a single power-law representing the combination of thermal and nonthermal emission. Distributions of the full- ($\alpha_{3 - 33 GHz}$) and inter-band ($\alpha_{3 - 15 GHz}$ and $\alpha_{15 - 33 GHz}$) spectral slopes are given in Figure 4. The median spectral indices we measure from $3 -15$, $15 - 33$, and $3 - 33$ GHz are $-0.58 \pm 0.11$, $-0.27 \pm 0.23$, and $-0.51 \pm 0.13$, respectively.  This is consistent with the expectation for star-forming regions where the lower frequencies are predominately synchrotron-dominated, and that at higher frequencies the overall radio SED begins to flatten as the contribution of thermal emission increases (Condon 1992; Leroy et al. 2011; Murphy et al. 2013). Indeed a two-sided Kolmogorov-Smirnoff (KS) test yields a probability of $\leq 10\%$ that $\alpha_{3 - 15 GHz}$ and $\alpha_{15 - 33 GHz}$ are drawn from the same distribution.

By comparison, results from modeling the integrated radio SED's of U/LIRGs, which are dominated by their nuclear emission, show that the radio spectrum of many LIRGs remain steep even at high frequencies (Clemens et al. 2008, 2010; Barcos-Munoz et al. 2017; Tisanic et al. 2018). Therefore, we conclude that extranuclear star-forming regions in LIRGs have distinctly different radio spectral shapes, and show significantly less system-to-system variation relative to the integrated properties of local U/LIRGs.

This interpretation can be complicated by the fact that low-frequency radio emission traces cosmic rays, which potentially diffuse out of the area covered by our photometric apertures used as they lose energy. Using far-infrared and 22 cm radio emission maps for a sample of 29 nearby spiral galaxies, Murphy et al. (2006) reported a correlation between the typical CR electron propagation length and the disk-averaged star formation rate, where CR electron propagation is found to decrease with increasing star formation activity (Equation 5). Murphy et al. (2012b) used analogous observations of the LMC and 30 Doradus to derive a CR electron propagation length of $\sim100-140$ pc (corresponding to a $\tau_{cool} \sim 1$x$10^{5}$ yr). These values are consistent with the empirical trend describing spiral galaxies, extrapolated to environments with nearly an order of magnitude higher star-formation rate surface density (log$(\Sigma_{SFR}) \sim -1$). Using this scaling relation, the measured star formation rate surface densities for the regions in our sample ($-1.5 <$log$(\Sigma_{SFR})< -0.5$: See the following Section), suggest an average CR electron propagation length of $\sim100-200$ pc; this is still several times smaller than the physical scale of our aperture in the closest galaxy in the sample (NGC 0838). Thus we do not expect any missing synchrotron emission on the scales of our photometric apertures due to cosmic ray diffusion.


\subsection{Thermal Fractions at 33 GHz}

As a pilot study, Murphy et al. (2012) used the Westerbork Synthesis Radio Telescope (WSRT) and Green Bank Telescope (GBT) to construct $1.4-33$ GHz SEDs for 50 normal star-forming galaxy nuclei and extranuclear star-forming complexes at a matched resolution of $\sim 25"$. They found evidence that the median thermal fraction at 33 GHz ($f_{th}$) was $\sim 80-90\%$ on physical scales of $\leq 0.5$ kpc, with the fraction decreasing to $\sim 60\%$ as the projected size of the photometric aperture increased to $\sim 1$ kpc. This study served as the basis for the Star Formation in Radio Survey (SFRS: Murphy et al. 2018a), which is a VLA campaign designed to extract the same multi-frequency band photometry as GOALS for 118 galaxy nuclei and extranuclear star-forming regions in a sample of 56 nearby lower-luminosity galaxies at a matched-resolution of $\sim 2"$. By comparing the 33 GHz radio emission to H$\alpha$ and 24$\mu$m observations of 225 discrete star-forming regions, Murphy et al. (2018a) demonstrates that the striking morphological similarities between these tracers on $50-100$ pc scales requires the emission from all three to be powered by the same source, namely massive star formation. The complete $3-33$ GHz SFRS data and associated analysis will be presented in a forthcoming paper.

\begin{figure}
\centering
\includegraphics[scale=0.43]{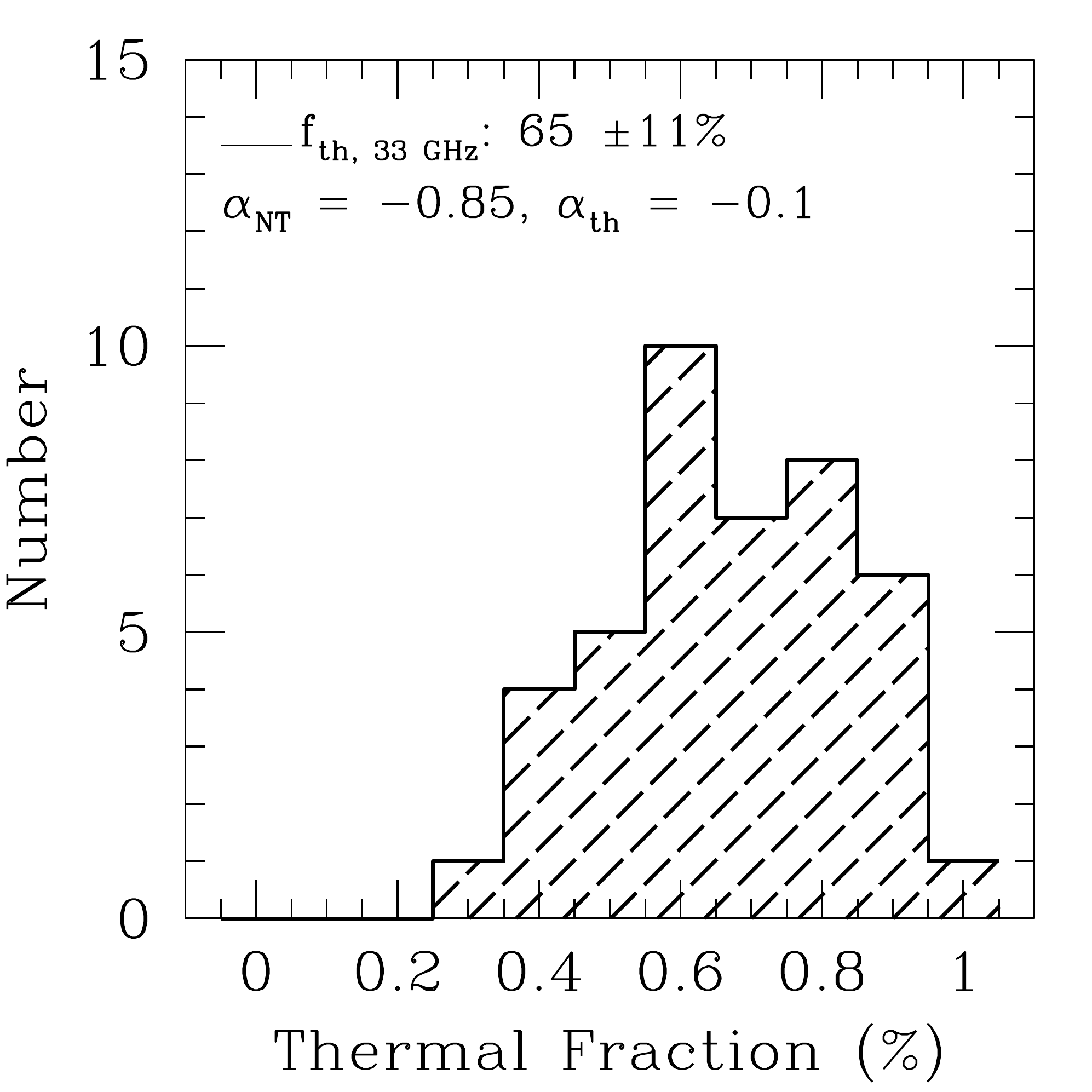}
\caption{The measured 33 GHz thermal fraction values for all extranuclear regions identified in the GOALS sample.~The median fraction is $65 \pm 11\%$ for the empirical calibration presented in Equation 2. This value is in good agreement with estimates of the thermal fraction made for star-forming regions in normal star-forming galaxies in the SFRS on the same physical scales (Murphy et al. 2012a, 2018a).}
\end{figure}

In contrast, nonthermal synchrotron has been observed to be the dominant contributor to the global 33 GHz emission (i.e. $f_{th} \leq 50 \%$ for $10^{11} < L_{IR} < 10^{12}$) of local U/LIRGs (Barcos-Mu\~noz et al. 2017). A possible explanation as to why LIRGs may have significantly lower thermal fractions relative to normal star-forming galaxies is that the absorption of a large fraction of ionizing stellar photons by dust grains, densely concentrated in starburst regions, suppresses the production of thermal radio emission relative to what is seen in the more diffuse star-forming regions in normal galaxies. Here we extend this investigation to isolated extranuclear regions in the disks of our LIRG sample.

To calculate the ratio of thermal/nonthermal emission at 33 GHz we use the spectral index, measured in Section 4.1, from $3 - 15$ GHz ($\alpha_{3 -15 GHz}$) to set the lower-limit on the nonthermal spectral index ($\alpha_{NT}$) such that  $\alpha_{NT} = -0.85$ if $\alpha_{3 -15 GHz} \geq -0.85$, and $\alpha_{NT} = \alpha_{3-15 GHz}  -  0.1$ if $\alpha_{3 -15 GHz} < -0.85$. This latter equation accounts for the fact that the measured 3 - 15 GHz radio spectral slope contains contributions from both non-thermal and thermal free-free emission components, and ultimately represents a lower-limit on the true thermal fraction at 33 GHz. Importantly, only 3/42 regions have $\alpha_{3-15 GHz} < -0.85$, and thus we adopt $\alpha_{NT} = -0.85$ for the majority of the regions in our sample. Further, the removal of these three regions does not affect the median of the measured thermal fraction distribution, and are thus not biasing our results in any way. Finally, we assume the same power-law exponent for the thermal emission ($\sim -0.1$), and use the fitted slope from $3-33$ GHz to set the overall SED shape. Then, using the prescription in Murphy et al. (2012), we can calculate the thermal fraction at 33 GHz such that,

\begin{equation} 
f_{T}^{\nu_{1}} = \frac{(\frac{\nu_{2}}{\nu_{1}})^{-\alpha} - (\frac{\nu_{2}}{\nu_{1}})^{-\alpha^{NT}}}{(\frac{\nu_{2}}{\nu_{1}})^{-0.1} - (\frac{\nu_{2}}{\nu_{1}})^{-\alpha^{NT}}}
\end{equation}

\noindent where $\nu_{1}$ is the target frequency, $\alpha$ is the observed slope from $\nu_{1}$ to $\nu_{2}$, and $\alpha^{NT}$ is the nonthermal spectral index. These values are given in Appendix Table A.1.


In Figure 5 we show the resulting thermal fractions of the star-forming regions identified in our sample using the shape of the radio SED as determined in Equation 2. We find that the median thermal fraction is $65\%$ at 33 GHz on $\sim$ kpc-scales in the extranuclear regions of LIRGs. This value is in good agreement with estimates of the thermal fractions for star-forming regions in normal star-forming galaxies on the same physical scales (Murphy et al. 2012a). The results of Figures 4 and 5 strongly suggest that while extranuclear star-forming regions in LIRGs have a non-negligible contribution from non-thermal synchrotron emission, these regions are much more heavily-dominated by thermal free-free emission relative to the resolved nuclei of local ULIRGs, which in the case of  Arp 220 can be as low as $\sim 20\%$ on $\sim 50$ pc scales, and high-redshift star-forming galaxies in the VLA-COSMOS survey (Barcos-Mu\~noz et al. 2015; Tisanic et al. 2018).

\subsection{Star Formation Rates}

We have shown in the previous two sections that one can use 33 GHz emission to reliably trace the current star formation activity of kpc-sized regions in the disks of both normal and extreme star-forming galaxies in the local Universe. Therefore, we assume that the thermal and nonthermal spectral indices do not vary significantly across the individual extranuclear regions in our sample, and that any differences in the observed SED are due to the relative contribution of each emission mechanism. This allows us to use Equation 10 presented in Murphy et al. (2012) to convert the measured 33 GHz luminosity ($L_{33 GHz}$) into the current SFR within our apertures, where the assumed thermal and nonthermal power-law indices are $-0.1$ and $-0.85$ respectively, and the electron temperature of the gas is $10^{4}$K. In Figure 6 we show that the measured range of SFRs for our sample is $\sim 0.05 - 7.5 M_{\odot}$yr$^{-1}$, with a clear peak at $\sim 1 M_{\odot}$yr$^{-1}$.


\begin{figure}
\centering
\includegraphics[scale=0.43]{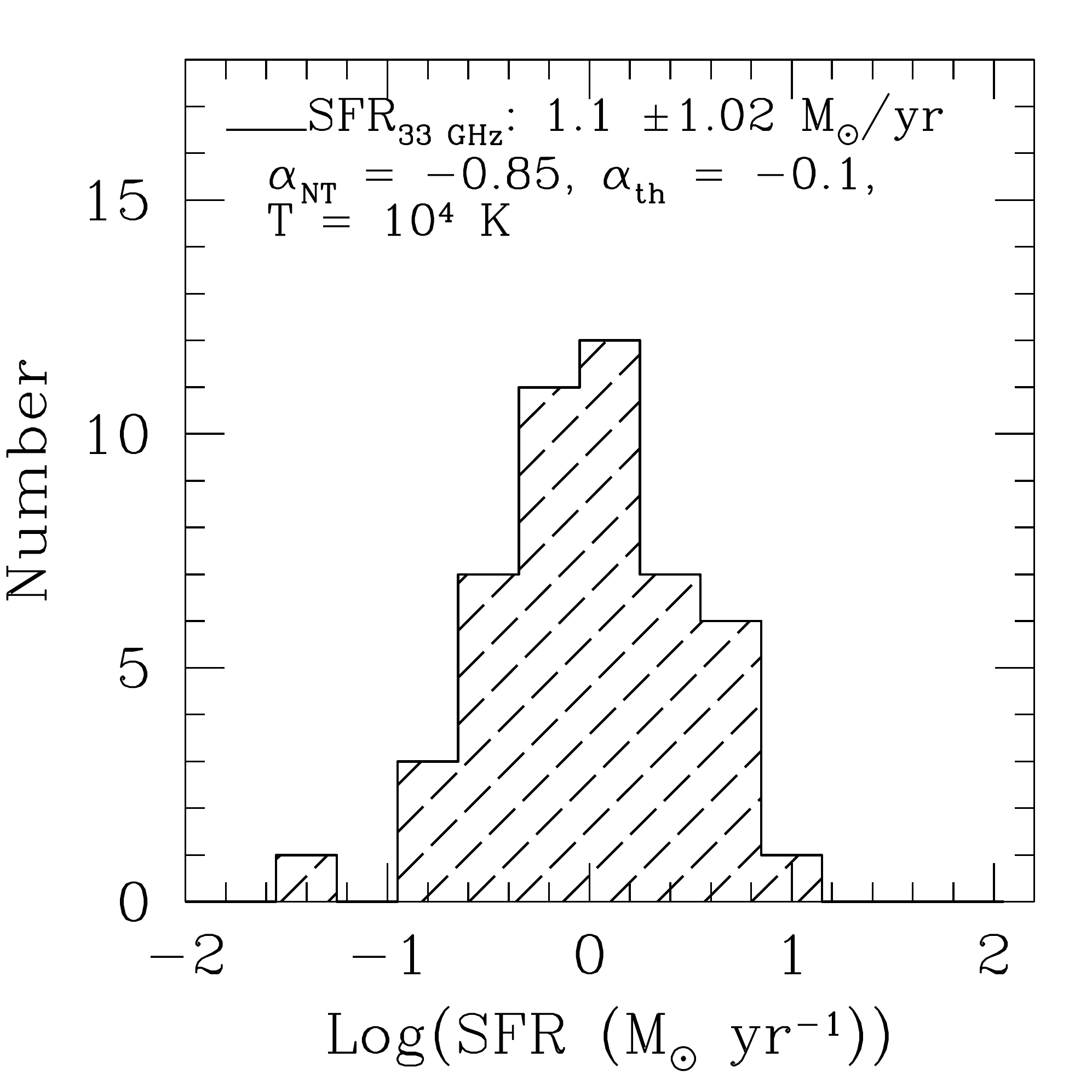}
\caption{The distribution of star-formation rates, as inferred from the measured 33 GHz luminosity ($L_{33 GHz}$), is shown along with the spectral indices assumed for the thermal and nonthermal emission components at 33 GHz. The SFR distribution for our sample shows a clear peak at $\sim 1 M_{\odot}$yr$^{-1}$, and lies beyond the upper-end of the range observed for star-forming regions in normal star-forming galaxies in the local Universe (Smith et al. 2016).}
\end{figure}



Smith et al. (2016) used GALEX near-UV (NUV) and far-UV (FUV) maps, along with Spitzer IRAC (Ch $1-4$), MIPS (24 $\mu$m), and archival H$\alpha$ images to estimate the extinction-corrected SFRs, in 1 kpc apertures, for nearly 700 star-forming regions in 46 interacting and non-interacting galaxy pairs. Importantly, they classify regions in their interacting galaxy sample into "inner-disk", "tidal", and "nuclear" regions, whereas, regions identified in the spiral galaxy sample were classified as either "disk" or "nuclei." They find that the distribution of star-formation rates for both "tidal" and normal galaxy "disk" regions range from $10^{-4} <$ SFR $< 10^{-1} M_{\odot}$ yr$^{-1}$. In Figure 6, we see that all but one of our regions, as measured in the same $\sim1$ kpc apertures, lie beyond the upper-end of this range.

In a new study, Larson et al. (2019) identified over 750 extranuclear star-forming regions in 50 U/LIRGs in GOALS using Pa$\alpha$ and Pa$\beta$ line emission. These regions range in size from $\sim 50 - 500$ pc with SFRs as low as 0.001 $M_{\odot}$yr$^{-1}$ up to 10 $M_{\odot}$yr$^{-1}$, consistent with the high SFRs observed for the regions in our sample. These results confirm that a significant population of star-forming regions exist at much lower SFRs in LIRGs, and that the peak observed in our distribution represents the sensitivity of our current radio observations to regions in the outer-disks of these galaxies. Therefore, while we cannot draw conclusions for the entire population of star-forming regions seen in LIRGs, it is clear that the most luminous extranuclear star-forming regions, as identified in the radio, are not seen in large samples of normal \textit{and} interacting galaxies in the local Universe (Smith et al. 2016). This is consistent with numerical simulations which show tidal disturbances can trigger enhancements of the gas turbulence and pressure in the ISM throughout the disks of luminous galaxy mergers, which leads to larger fractions of dense gas, and thus more massive star-forming regions (e.g. Elmegreen et al. 1993; Hopkins et al. 2008; Struck \& Smith 2012; Kruijssen 2014). In Section 5 we will examine how the measured SFRs in each region compare to the IR and $M_{*}$ properties inferred from our near-IR observations.

\subsection{Model Age Fitting}

In this Section we estimate the age of the starburst in each region by examining how thermal and nonthermal affect the measured 3 - 33 GHz radio spectral slope of the starburst as it ages. Since thermal emission is only produced by the shortest-lived ($\leq 10$ Myr) massive stars, its presence in large amounts relative to synchrotron emission is indicative of very young star formation. This correlation can then be a method of determining approximate ages for the global star formation history of a galaxy. 

To quantify these different timescales, we use Starburst 99 (SB99) models with default inputs (solar metallicity and 2-component Kroupa Initial Mass Function), in order to estimate the ionizing photon rate ($Q_{H_{0}}$ s$^{-1}$) and supernova rate (N$_{SN}$ yr$^{-1}$) of a simple stellar population (SSP; Leitherer et al. 1999). To transform these quantities into a theoretical  3 - 33 GHz spectral index as a function of time we use Equations 5 and 8 in Murphy et al. (2012) respectively, to estimate the total, thermal, and nonthermal luminosity at 3 and 33 GHz (assuming the same values for the thermal and nonthermal spectral index as the previous sections). We stress that this model does not include losses due synchrotron, inverse Compton scattering, or free-free self-absorption. Instead the model is meant to illustrate the effect an aging stellar population has on the radio spectral slope of an isolated HII region. Further, given that the cooling timescale estimated for CR elections in Section 4.1 ($\sim 10^{5}$ yr) is equal to the time-step used in SB99, we expect synchrotron losses to have a negligible effect on the 3 - 33 GHz radio spectral slope. Rabidoux et al. (2014) used this simple framework to describe the 1.4-33 GHz spectral slope, and thus the average age of the global star-formation activity, in a sample of 27 nearby normal star-forming galaxies. With our sample, we are able to extend this analysis to individual star-forming regions in the disks of luminous starburst galaxies.

\begin{figure*}
\centering
     \includegraphics[scale=0.43]{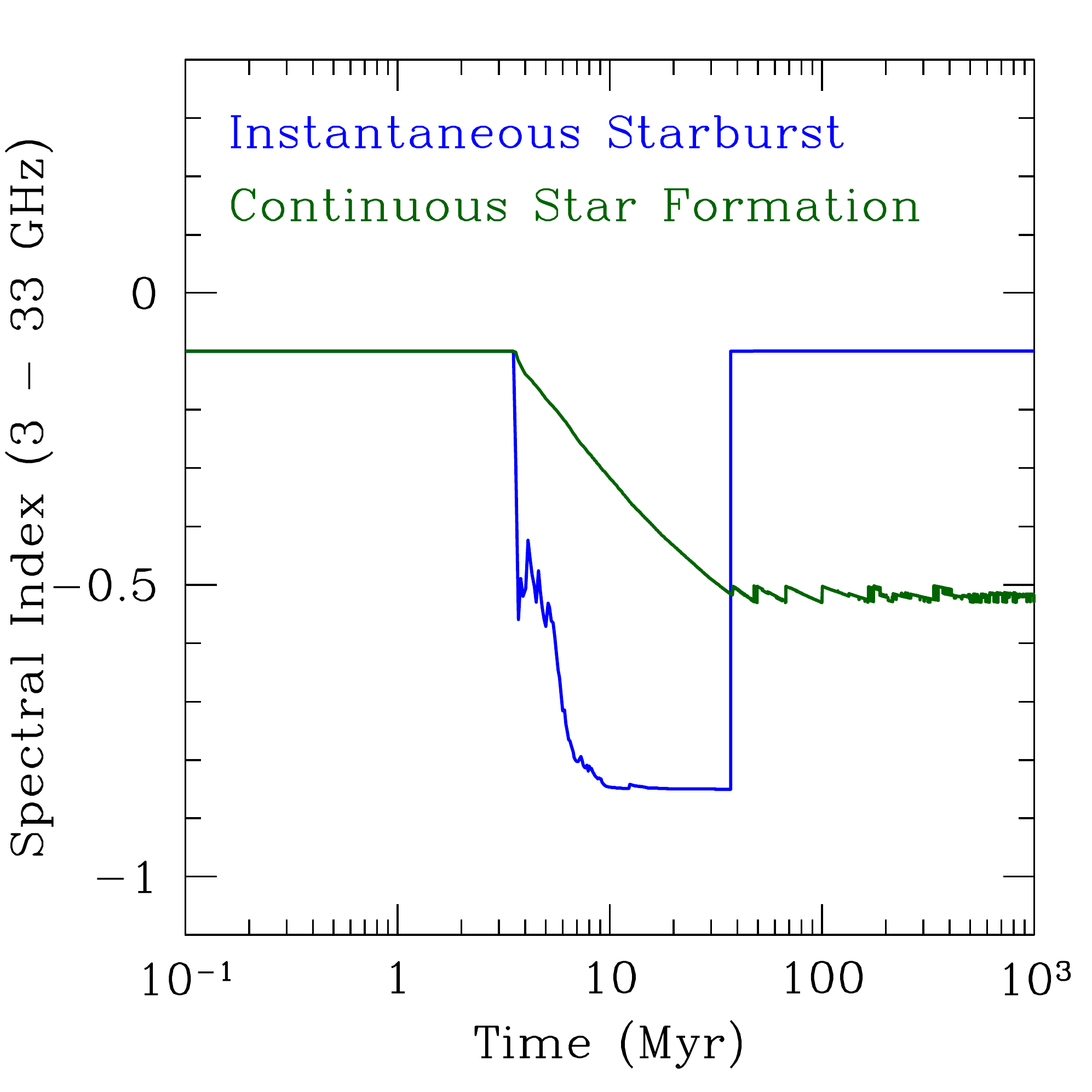}
     \includegraphics[scale=0.43]{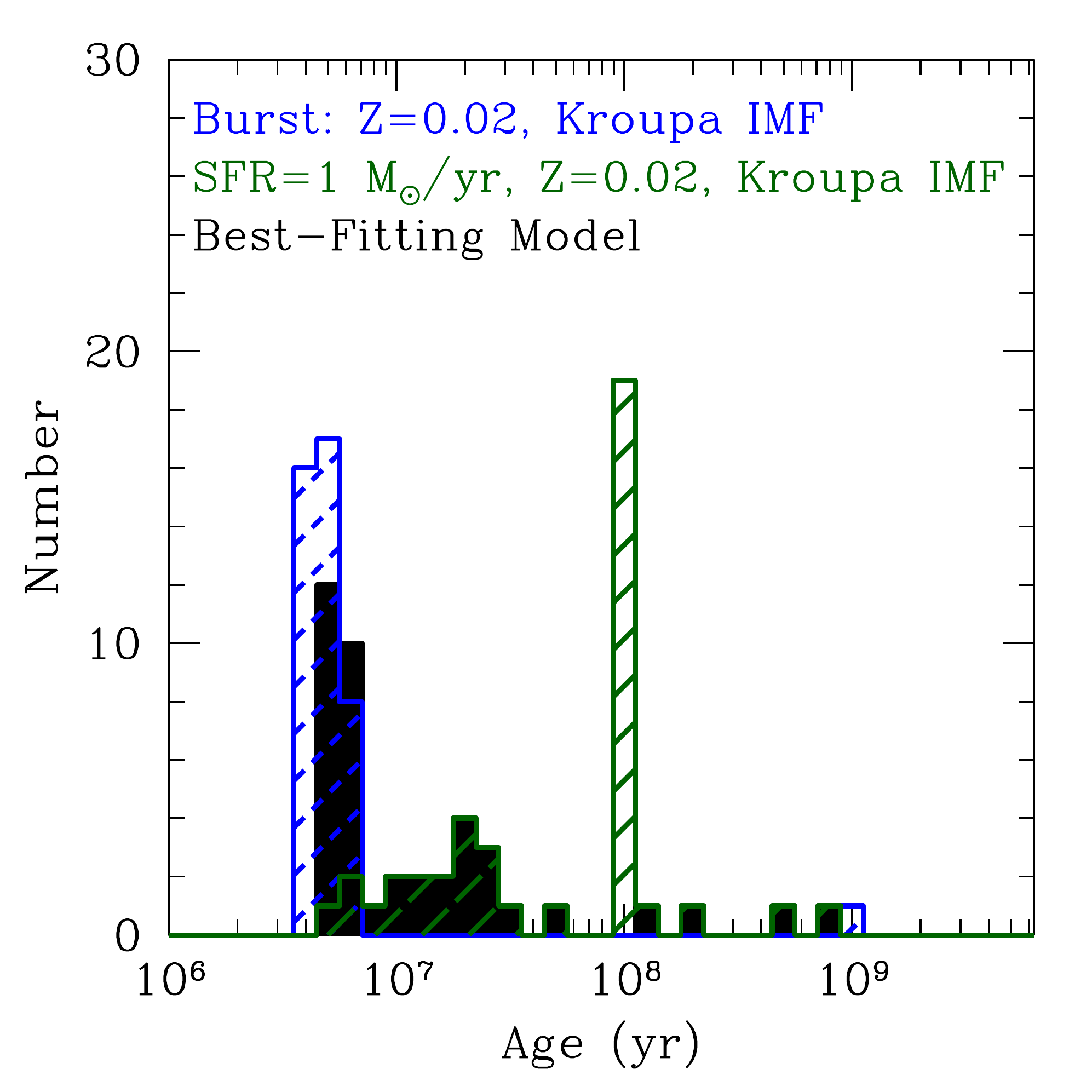}
\caption{Left panel: The two sub-panels show SB99 models of both an instantaneous burst and continuous SFH, using standard Kroupa IMF, and solar metallicities. We perform a $\chi^{2}$-minimization to the observed $3-33$ GHz spectral index of each region. Right panel: The distribution of model ages for both types of SFH (blue and green) and the best-fitting model in each case (black). It is clear that there exists two populations of regions: Those younger than $t \sim 10^{7}$ yr, which are best-modeled by an instantaneous burst, and those older than $t \sim 10^{7}$ yr, which are best-modeled by a continuous SFH.}
\end{figure*}

To take into account the fact that a single instantaneous starburst may not be representative at the physical scales we are probing ($\sim$1 kpc), we also include SB99 models with a continuous star formation history (SFH) of $SFR = 1M_{\odot}$yr$^{-1}$ using the same metallicity and IMF input as in the instantaneous burst model. One can see that in the continuous SFR model, there are no large jumps in the 3-33 GHz spectral index. Instead, the model transitions smoothly from a very shallow (thermal dominated) spectral slope to an intermediate spectral slope. This model predicts that a spectral index of -0.5 would be expected for a kpc-sized region that has been actively forming stars for $\sim 10$ Myr. This is consistent with the median spectral index of -0.51 measured for the star-forming regions in our sample.

Using both models, we then perform a $\chi^{2}$ minimization to the observed spectral index for each extranuclear region in the sample. The right panel of Figure 7 shows the distribution of fitted-ages considering only the instantaneous burst model in blue, only the continuous model in green, and whichever model better-fits the observed spectral index of each region in black. Overall, the estimated median age of our star-forming regions is $\sim 10$ Myr, which agrees with the age distributions derived for a large sample of young massive star clusters in a sample of local U/LIRGs in GOALS (Linden et al. 2017). By comparing each model individually, we see that 45$\%$ of regions in our sample are best modeled by an instantaneous burst, which indicates that these regions are very young. However, the majority of the star-forming regions are best fit with a continuous SFR model with ages between $10^{7} - 10^{7.5}$ yr, and even some regions which are relatively old ($t > 10^{8}$ yr). Finally, when examining the ages of these regions as a function of merger stage we do not see evidence that the oldest regions are observed exclusively in the latest-stage mergers. This indicates that while the nuclear starburst activity dominates as the merger progresses, prodigious star formation still occurs in the outer-disks of these systems.

\section{Discussion}

For the following analysis, we first create a sub-sample of the 48 star-forming regions, which have $3-33$ GHz spectral indices which span a parameter space that can be reliably modeled with one of our two SB99 models ($-1.2 < \alpha_{3 -33 GHz} < 0.0$). This is to ensure that any differences in the slopes observed are not due to regions that are faint in one of the three radio bands, which could affect both the measured spectral index, and the inferred low-frequency radio luminosity. Particularly, if the region is faint at 33 GHz the assumption that the region is fully sampling the IMF, a key detail which underpins all SFR calibrations, might break down. In total we retain 42/48 regions identified in the initial sample, and in fact the six regions we remove have the lowest $S/N$ ratios across all three bands we identified in the initial candidate selection.

\subsection{The Infrared -- Radio Correlation}

The far-infrared -- radio correlation (Helou et al.\ 1985) is an empirical relationship which holds remarkably well for galaxies spanning a wide range in mass and luminosity (Yun et al. 2001). At centimeter wavelengths, the radio continuum is dominated by synchrotron emission, which decays on timescales of $\sim 100$ Myr for pseudo-continuous star formation in galaxies (Condon et al.\ 1992). The infrared (IR) traces the peak of the dust emission, which is a proxy for recent star formation in a starburst galaxy. For a fixed initial mass function (IMF) and star formation history, the production of cosmic rays is roughly equal to the rate of dust-heating from UV photons produced by young stars, such that the IR-radio correlation holds in both normal and extreme star-forming galaxies in the local Universe (Lisenfeld et al.\ 1996). The physical explanation for the tightness of this correlation has long been debated, given the fact that the methods which produce each emission mechanism have timescales which differ by an order of magnitude. 

However, when individual star-forming regions on smaller spatial scales ($\sim$ hundreds of parsecs) are examined in both normal and extreme star-forming galaxies, this correlation can break down, and is sensitive to these various timescales and local SFH, CR propagation, and metallicity of the galaxy (Murphy et al. 2006). Further, it has been shown with lower-resolution data that ongoing mergers, whose progenitors still share a common envelope, may also exhibit excess radio emission from bridges and tidal tails that is unassociated with the current star-formation activity. This scenario may also explain the seemingly low far-infrared/radio ratios measured for many high-z submillimeter galaxies, a number of which are merger-driven starbursts (Murphy et al. 2013). Here we aim to test whether the infrared -- radio correlation will hold on kpc-scales in the extended disks of LIRGs at various stages along the merger sequence, and determine to what level we see evidence for excess non-thermal emission relative to the inferred far-infrared luminosity within individual star-forming regions. 


In order to make accurate measurements of infrared luminosity, we require observations at matched resolution to our VLA images, for which MIPS and Herschel data are not sufficient. We therefore used the IRAC Ch4 $8\mu$m flux as a proxy for the total-IR ($L_{1-1000 \mu m}$) by assuming a fixed IR8 ratio ($L_{TIR}/L_{8 \mu m}$). Elbaz et al. (2011) used observations of IRAS selected galaxies, including the full GOALS sample, to determine that the global ratio of total IR luminosity to rest-frame 8$\mu$m luminosity follows a distribution centered on IR8 = 4, thus defining an IR main sequence for star-forming galaxies independent of redshift. This study was limited by the fact that the galaxies in their sample were not classified into AGN- and SF-dominated systems. Wu et al. (2010) and Magdis et al. (2013) used the 5MUSES sample of galaxies, which builds on the Elbaz et al. (2011) sample by including spectral diagnostics from Spitzer and Herschel respectively, to isolate galaxies which are dominated by star-formation. Both studies concluded that the IR8 ratio was larger (by up to a factor of 2), than what Elbaz et al. (2011) found for the complete galaxy sample. Following these studies, we used the global $L_{8\mu m}$ and $L_{IR}$ photometry presented in Chu et al. (2017) and Mazzarella et al. (2019 in prep) to derive a median IR8 ratio of $8.1 \pm 2$ for the 25 galaxies in our equatorial sub-sample. This is in good agreement with the Wu et al. (2010) and Magdis et al. (2013) calibrations, and consistent with the notion that once AGN-dominated galaxies are removed from the global IR-bright galaxy population the ratio is significantly enhanced in pure starburst-dominated galaxies. 

Using the measured $3-15$ GHz radio spectral slope ($\alpha_{3-15 GHz}$) we extrapolate the observed 3 GHz flux density of each region to 1.4 GHz and measure the q-ratio ($q_{TIR}$) defined as the logarithmic ratio of the total infrared to radio flux density for each region. While this measurement differs from the traditional $q_{FIR}$ analysis discussed above, it allows us to make direct comparisons with a recent calibration of the global TIR-radio correlation observed for a large sample of normal star-forming galaxies in the local Universe (Bell et al. 2003). Overall, we find that the median $q_{TIR}$ derived for kpc-sized regions in our LIRG sample ($2.7 \pm 0.34$) is consistent with the Bell et al. (2003) calibration ($q_{TIR} = 2.64$). While the uncertainty in our IR8 calibration limits the robustness of this result, we do not see strong evidence for regions with a significant excess nonthermal emission associated with tidal bridges and tails. With future ALMA and pre-approved JWST/MIRI programs we will further investigate the physical origin of the infrared -- radio correlation by directly measuring the total infrared luminosity, dust, and gas masses of individual star-forming clumps identified in GOALS galaxies.


\subsection{The Star-Formation Main Sequence}

The relationship between the star-formation rate (SFR) and the observed stellar mass ($M_{*}$) in galaxies has been extensively investigated over the past decade as means for understanding the evolution of galaxies (e.g., Noeske et al. 2007; Daddi et al. 2007; Elbaz et al. 2011). From these studies it is clear that there are two main modes of star formation that are known to control the growth of galaxies: a relatively steady rate, which defines the star formation rate - stellar mass main sequence (SFMS), and a starburst mode above this sequence. Further, homogeneous collections of the integrated SFMS of galaxies across large ranges in redshift (e.g. Speagle et al. 2014; Kurczynski et al. 2016) show that while the slope remains relatively constant, the fitted zero-point of the relation appears to increase at higher ($z > 1$) redshifts ($SFR \propto M_{*}^{0.5}$: Speagle et al. 2014), indicating a more significant contribution from starburst galaxies at earlier times. 

More recently, several studies (e.g., Wuyts et al. 2013; Cano-Diaz et al. 2016; Maragkoudakis et al. 2016; Medling et al. 2018) have provided evidence that MS-like correlations are also present at sub-galactic scales in a wide variety of galactic environments, by comparing the SFR surface densities ($\Sigma_{SFR}$) with stellar-mass surface densities ($\Sigma_{M_{*}}$) for individual sub-galactic regions $\sim$ 1kpc or larger. Thus far observations at high-redshift suggest this correlation has a slope close to unity (Wuyts et al. 2013), whereas at lower redshifts linearity and sub-linearity have been reported (Cano-Diaz et al. 2016; Medling et al. 2018). Here we look to test which of these two modes of star-formation best describes the extranuclear star-forming regions identified in the equatorial GOALS sample, and whether or not a MS-like correlation exists for the most luminous regions in LIRGs.

\begin{figure*}
\centering
     \includegraphics[scale=0.43]{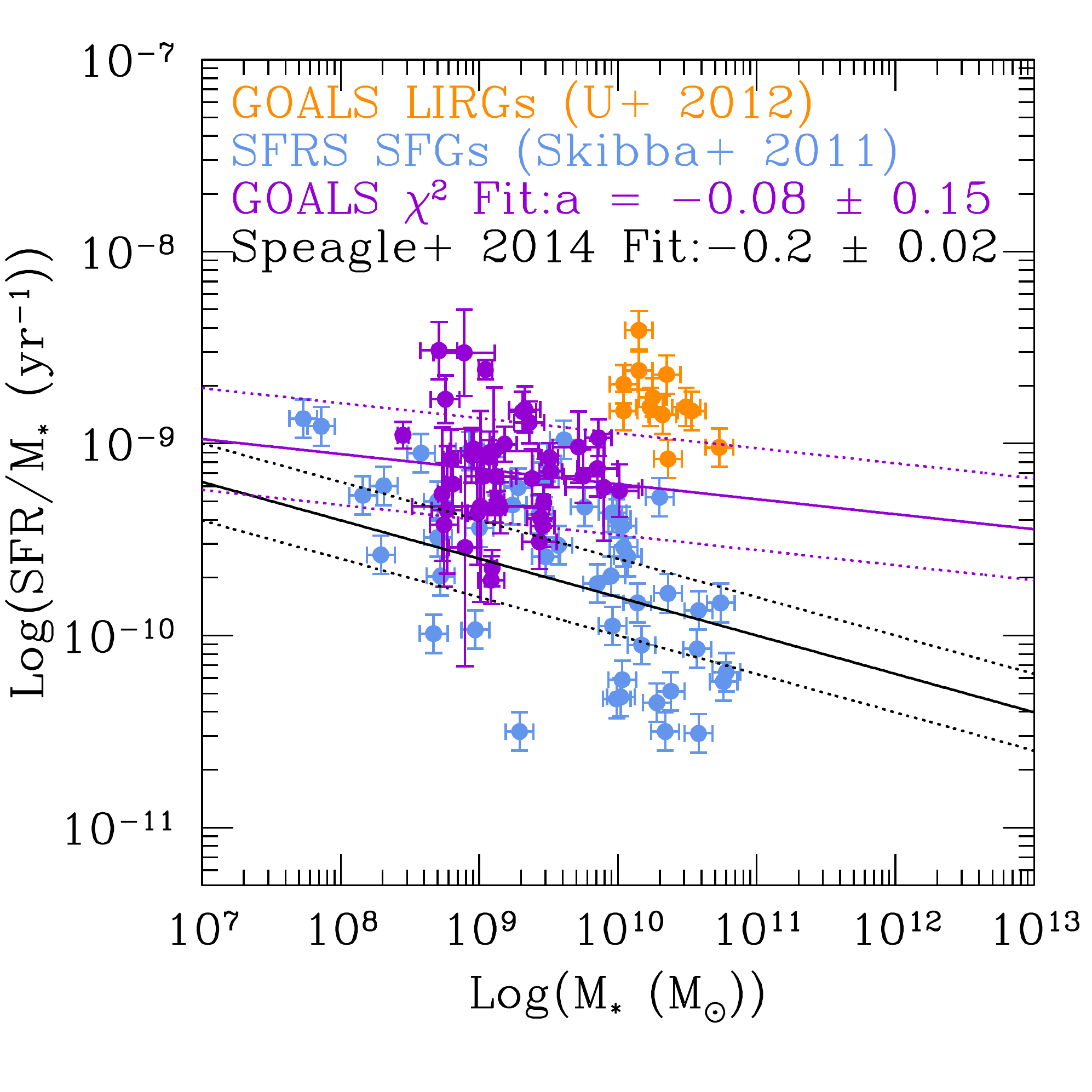}
     \includegraphics[scale=0.43]{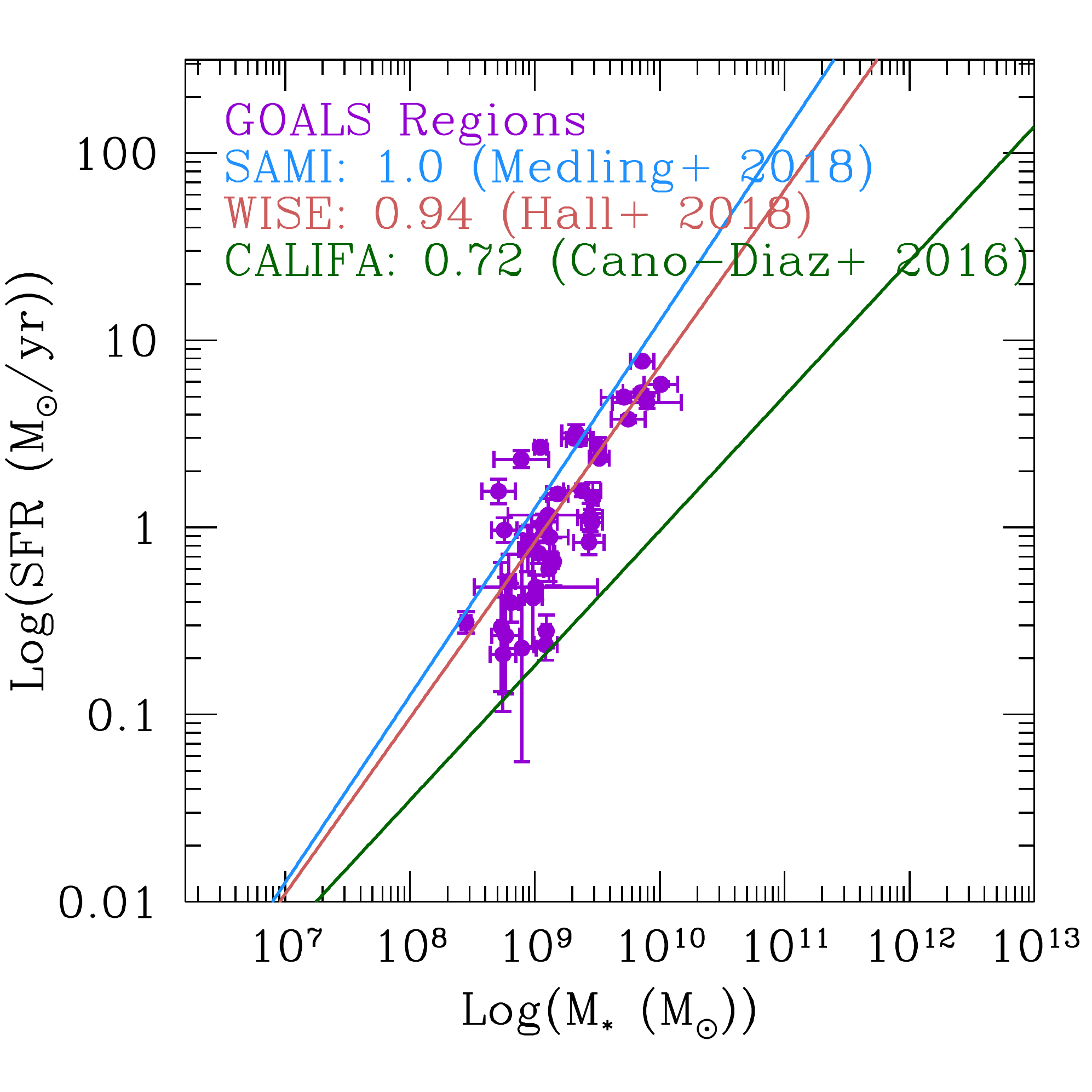}
\caption{Left panel: The sSFR distribution as a function of stellar-mass showing that the global fit from Speagle et al. (2014) is not an appropriate calibration for both the galaxy-integrated (GOALS: orange) and extranuclear star-forming regions in local LIRGs (purple). Further, the specific star formation rate of the extranuclear regions themselves lie at the upper-end of the relation for normal star-forming galaxies in the local Universe (SFRS: blue, Skibba et al. 2011). Right panel: The star-formation main sequence ($SFR-M_{*}$). In purple we show the results for the extranuclear star-forming regions identified in GOALS. The blue, dark green, and red lines show the resolved galaxy main-sequence, normalized to the GOALS fit presented in the left panel, for the SAMI, CALIFA, and WISE surveys respectively.}
\end{figure*}

In Figure 8 we compare the star-formation rates and stellar masses of our extranuclear regions with both integrated galaxy properties (left panel) and correlations found for the sub-galactic main sequence (right panel). In the left panel the orange points are globally-integrated measurements from our sub-sample of 25 galaxies taken from U et al. (2012), and in purple are our extranuclear star-forming regions identified with the VLA. The integrated stellar mass measurements from U et al. (2012) used the observed H-band luminosity and a Chabrier IMF, whereas for our SFR calculations and the IRAC-$M_{*}$ conversion we utilize a Kroupa and Chabrier IMF respectively. However, the differences in the integrated total mass are small compared to our measured uncertainties, and therefore we do not expect to introduce any systematic biases when comparing the two datasets using slightly different IMFs. The light blue points are integrated measurements of galaxies in the SFRS as a reference sample of normal star-forming galaxies (Skibba et al. 2011; Murphy et al 2018a). The solid black line shows the main-sequence as defined by a large homogeneous collection of local spirals taken primarily from SINGS (Speagle et al. 2014). We can see that the galaxies in the SFRS generally follow the local MS calibration, and that the massive extranuclear regions identified in our LIRG sample are consistently above it.

If the integrated measurements of starburst-dominated galaxies were simply the sum of the individual star-forming regions identified within them, then the integrated LIRGs and the extranuclear regions should be well described by a single linear fit. Shown in purple is the fit to only the extranuclear regions, and we can see that the integrated LIRG measurements are systematically offset by $\sim0.2$ dex in sSFR. By comparing the median values of sSFR for both the integrated GOALS LIRGs and our individual star-forming regions we find that they make up on average $16.5\%$ of the current star formation activity ($f_{SFR} = SFR_{region}/SFR_{galaxy}$) in their host galaxies. This is consistent with a recent suite of 75 hydrodynamic simulations of major galaxy mergers ($M_{rat} \sim 2.5:1$), which show a median $f_{SFR} \sim 13\%$ in regions from 1 - 10 kpc away from nucleus over a broad range of interaction geometries (Moreno et al. 2015). Further, it is clear that the fit to the local galaxy reference sample is steeper than the relation found for the most luminous GOALS regions identified. Ultimately, this suggests that while the integrated properties of the starburst-dominated LIRGs are driven by the central nuclear starburst, extranuclear regions in LIRGs have elevated sSFRs even relative to normal star-forming galaxies.



In the right panel of Figure 8 we show a plot of the star-formation - stellar mass plane for our identified regions with fits to the sub-galactic main sequence overlaid.~By normalizing the zero-points of each fit we can test which correlation most accurately represents our star-forming regions, and to what degree the linear or sub-linear relationships found for integrated galaxies hold on kpc-scales in LIRGs. We see clearly that the correlation found for the Calar Alto Legacy Integral Field Area (CALIFA) survey of galaxies is shallower than our distribution of star-forming regions. This discrepancy is likely due to the range of morphological types included in the CALIFA fit, with regions from early-type galaxies systematically flattening the slope of the correlation (Cano-Diaz et al. 2016). Indeed, when the fit is restricted to only late-type galaxies the correlation closely follows the distribution of star-forming regions observed in our VLA sample (Hall et al. 2018; Medling et al. 2018). This confirms that there is a sub-galactic main sequence of star-formation present in LIRGs with both individual star-forming regions and the globally-integrated galaxy measurements, which lie above the locally-calibrated SFRMS.


\subsection{Spatially Coincident Massive Star Clusters}

\begin{figure*}
\centering
\includegraphics[scale=0.38]{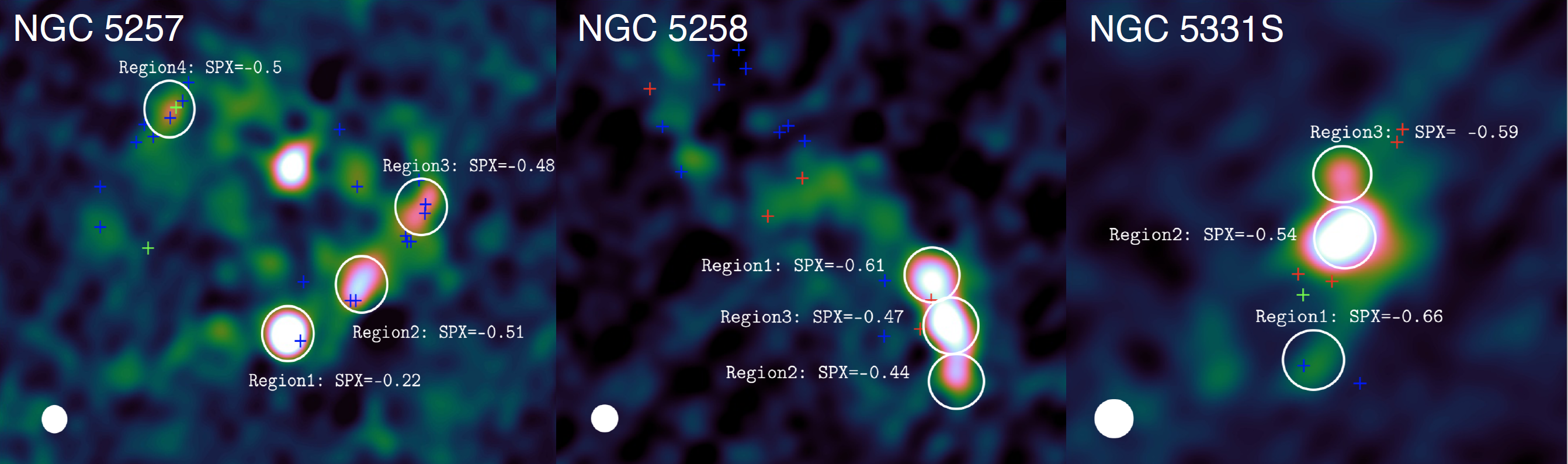}
\caption{The 33 GHz continuum images for NGC 5257 (Left), NGC 5258 (Middle), and NGC 5331S (Right) are shown overlaid with our photometric apertures, the measured $3 - 33$ GHz radio spectral index, and the locations of super star cluster identified in Linden et al (2017). The colors of the markers in all panels indicate their age, with $t \leq 10^{7}$ yr shown in blue, $ 10^{7} < t < 10^{8}$ yr shown in green, and $t \geq 10^{8}$ yr shown in red.}
\end{figure*}

Radio continuum emission has been used as an effective way to identify ultra-young ($1 - 3$ Myr) massive star clusters still deeply embedded in their natal birth material (Turner et al. 2000; Johnson et al. 2001; Johnson et al. 2003). Our results from the initial SFRS sample (Murphy et al. 2018a) reveal both purely thermal (and thus very young) sources, as well as sources which have higher nonthermal fractions at 33 GHz. In general, we expect to find that these latter radio sources are associated with regions that contain multiple star clusters visible at optical wavelengths (Evans et al. 2008; Inami et al. 2010; Modica et al. 2012; Mazzarella, et al. 2012; Mulia et al. 2015). Four galaxies in our sample, NGC 5257/8 and NGC 5331N/S, have been observed with HST as part of a larger program to search for young, UV-bright, massive star clusters in LIRGs (Linden et al. 2017). In this subsection, we compare the mean spectral index of each region, which tracks the relative fraction of young ($\sim$ 5 Myr) and old ($\geq 50$ Myr) star-formation to the median cluster age, which tracks the $\sim 1 - 100$ Myr SFH of the region within our photometric apertures.

At the resolution of our matched-VLA observations, it is unclear if these extranuclear regions are powered primarily by a single massive star cluster, or several lower-mass clusters tightly packed within a group. In Figure 9 we show the star clusters identified in these galaxies, color-coded by their modeled ages (i.e., blue $t \leq 10^{7}$ yr, green $10^{7} < t < 10^{8}$ yr, and red $t \geq 10^{8}$ yr). In only one case, NGC 5257, do we have significant overlap with regions identified in the radio for which a meaningful comparison of cluster ages, masses, and extinctions to the radio spectral slope can be done. We find that the region in NGC 5257 (left panel), which is the most luminous and shows the flattest $3 - 33$ GHz spectral index (-0.22), is associated with a single, young ($\sim 4$ Myr), massive ($M_{cl} \sim 10^{7} M_{\odot}$) star cluster, whereas regions with steeper radio spectral indices are coincident with several star clusters whose median age is slightly older ($\sim 10$ Myr) and mass significantly smaller ($\sim 10^{5.5} M_{\odot}$: Linden et al. 2017).

Overall we do see evidence, albeit in one system, that the luminosity and spectral index of the radio continuum measured at kpc-scale resolution is able to roughly track variations in the median age and mass of the spatially-coincident super star clusters identified in the UV/optical. A more thorough discussion of massive star clusters and their relationship to star-forming regions identified at 33 GHz  will be discussed in a forthcoming paper, which will be a comparison of the cluster mass functions, and luminosity distributions of young massive star clusters using HST data available for the SFRS sample of galaxies.



\section{Summary}

We have presented the first results of a high-resolution VLA survey for 25 luminous infrared galaxies (LIRGs) in the Great Observatories All-Sky LIRG Survey (GOALS). Radio emission provides a critical, optically-thin view on the massive star formation activity within deeply embedded HII regions, and it tracks nonthermal emission from relativistic cosmic rays associated with recent supernova in galaxies. We have extracted luminosities, spectral indices, star-formation rate (SFRs), thermal fractions ($f_{th}$), ages, and stellar masses for a total of 42 individual extranuclear star-forming regions identified as having de-projected galactocentric radii ($r_{G}$) which lie outside the 13.2$\mu$m core size of the galaxy measured in D\'iaz-Santos et al. (2010). These "extranuclear'" regions, allow us to cleanly examine the evolution of star-formation activity in LIRGs, free from possible contamination associated with an AGN. Our results indicate that:

\noindent
(1) The median $3-33$ GHz spectral index and thermal fraction at 33 GHz measured for the extranuclear regions identified in our VLA survey is $-0.51 \pm 0.13$ and $65 \pm 11\%$ respectively. These results suggest that on kpc-scales extranuclear star-forming regions in LIRGs have flatter radio spectral slopes, and are much more heavily-dominated by thermal free-free emission relative to the centers of local U/LIRGs. Further, the median $3-33$ GHz spectral index observed is consistent with models of continuous star-formation activity over a median lifetime of $\sim10$ Myr.

\noindent
(2) The median derived SFR of the extranuclear regions identified is $\sim 1 M_{\odot}$yr$^{-1}$. Despite the sensitivity of our observations to low-mass star-forming regions LIRGs, it is clear that the most luminous extranuclear star-forming clumps identified in our survey are not seen in large samples of normal \textit{or} interacting galaxies in the local Universe (Smith et al. 2016).


\noindent 
(3) The median $q_{TIR}$ derived for our extranuclear star-forming regions ($2.71 \pm 0.34$) is broadly consistent with the IR-radio correlation measured for normal and extreme star-forming galaxies in the local Universe (i.e. $q_{TIR} = 2.64$). This suggests that on kpc-scales in LIRGs we are sampling a representative volume of the ISM over a sufficiently long SFH so as to cause these regions to lie along the correlation. 

\noindent
(4) When we place our regions on the star-formation rate main sequence (SFMS), we find that they are not consistent with their host galaxies' globally-averaged specific star-formation rate (sSFR). This indicates that the nuclear starburst activity predominately drives LIRGs above the SFMS. 

With maps of the star-forming regions which energize LIRGs now in possession for the equatorial sample, the next step will be to obtain complementary high-resolution imaging and kinematics of the associated molecular gas, which fuels star formation and AGN activity in these extraordinary galaxies. The combined datasets would serve as a means to measure both the conditions under which star formation is most efficient, and energetic feedback on the ISM at scales that are inaccessible to extreme starbursts being studied at high-redshift.

\acknowledgements
S.T.L. was supported by the NRAO Grote Reber Dissertation Fellowship. The National Radio Astronomy Observatory is a facility of the National Science Foundation operated under cooperative agreement by Associated Universities, Inc. A.S.E., and Y.S. were supported by NSF grant AST 1816838. G.C.P. acknowledges support from the University of Florida. A.S.E. was also supported by the Taiwan, R.O.C. Ministry of Science and Technology grant MoST 102-2119-M-001-MY3. T.D-S. acknowledges support from ALMA-CONICYT project 31130005 and FONDECYT regular project 1151239.

Portions of this work were performed at the Aspen Center for Physics, which is supported by National Science Foundation grant PHY-1066293. This work was partially supported by a grant from the Simons Foundation. Finally, This research has made use of the NASA/IPAC Extragalactic Database (NED) which is operated by the Jet Propulsion Laboratory, California Institute of Technology, under contract with the National Aeronautics and Space Administration. This research has made use of the NASA/ IPAC Infrared Science Archive, which is operated by the Jet Propulsion Laboratory, California Institute of Technology, under contract with the National Aeronautics and Space Administration.


\appendix

In the following section we present both the derived (thermal fraction, SFR, age, etc.) and observed (IRAC and VLA) properties for 42 extragalactic star-forming regions identified in a sample of 25 galaxies from the Great Observatories All-Sky LIRG Survey. In Table A.1, the star formation rates at 33 GHz are calculated using Murphy et al. (2012a), and the Starburst99 model ages for both the continuous and instantaneous starburst are given for region. In Table A.2, all VLA and Spitzer IRAC photometry is given in mJy.

\begin{deluxetable}{lcccclcllll}
\tabletypesize{\footnotesize}
\tablecolumns{11}
\tablewidth{0pt}
\tablecaption {Derived Properties of Extranuclear Star-Forming Regions}
\tablehead{
\colhead{Region} & \colhead{RA} & \colhead{DEC} & \colhead{SFR\tablenotemark{a}} & \colhead{$f_{th, 33GHz}$} & \colhead{$L_{IR}$\tablenotemark{b}} & \colhead{$L_{1.4 GHz}$} & \colhead{$t_{inst}$\tablenotemark{c}} & \colhead{$t_{cont}$} & \colhead{$M_{*}$\tablenotemark{d}} & \colhead{sSFR\tablenotemark{e}}} \\
\startdata
MCG-02-01-051\_e1 & 00:18:49.81738 & -010.21.34.1910 & 0.88 & 0.33 & 36.17 & 20.50 & 9.00 & 6.34 & 8.96 & 7.94 \\
IC1623\_e1 & 01:07:46.69656 & -017.30.20.5717 & 1.09 & 0.56 & 36.50 & 21.26 & 6.68 & 8.00 & 9.11 & 14.15 \\
IC1623\_e2 & 01:07:46.74329 & -017.30.27.1956 & 2.17 & 0.58 & 36.81 & 21.59 & 6.70 & 8.00 & 8.89 & 13.80 \\
IC1623\_e3 & 01:07:47.07207 & -017.30.26.6457 & 1.47 & 0.75 & 36.75 & 21.41 & 6.57 & 8.00 & 8.71 & 8.82 \\
MCG-03-04-014\_e1 & 01:10:09.05601 & -016.51.12.7120 & 4.39 & 0.92 & 37.32 & 21.96 & 6.76 & 8.00 & 9.90 & 7.79 \\
NGC0838\_e1 & 02:09:38.30876 & -010.08.49.4349 & 0.45 & 0.42 & 36.52 & 20.68 & 6.61 & 7.10 & 9.01 & 1.82 \\
IC0214\_e3 & 02:14:05.02289 & +005.10.29.9168 & 3.01 & 0.20 & 36.48 & 21.68 & 6.59 & 8.61 & 9.33 & 8.57 \\
IC0214\_e2 & 02:14:05.30829 & +005.10.28.6774 & 2.55 & 0.48 & 36.80 & 21.67 & 6.70 & 8.00 & 9.51 & 27.93 \\
NGC0877\_e2 & 02:18:00.27225 & +014.32.27.0777 & 0.04 & 0.31 & 35.90 & 20.18 & 7.57 & 8.00 & 8.65 & 28.68 \\
NGC0877\_e1 & 02:18:00.42413 & +014.32.44.6577 & 0.29 & 0.37 & 35.75 & 20.26 & 9.00 & 6.59 & 8.45 & 19.40 \\
UGC02369\_e1 & 02:54:01.90424 & +014.58.11.5414 & 4.67 & 0.40 & 36.98 & 22.01 & 6.76 & 8.00 & 9.71 & 15.99 \\
CGCG465-012\_e2 & 03:54:16.15479 & +015.55.35.1392 & 0.80 & 0.42 & 36.24 & 21.13 & 6.68 & 8.00 & 8.96 & 7.99 \\
CGCG465-012\_e1 & 03:54:16.18669 & +015.55.47.5079 & 2.82 & 0.31 & 36.65 & 21.70 & 6.70 & 8.00 & 9.31 & 5.23 \\
CGCG465-012\_e3 & 03:54:16.56956 & +015.55.37.3463 & 0.48 & 0.40 & 36.00 & 20.86 & 6.64 & 7.51 & 8.79 & 6.32 \\
UGC02982\_e5 & 04:12:22.09337 & +005.32.59.1349 & 0.20 & 0.83 & 35.92 & 20.60 & 6.75 & 8.00 & 8.75 & 10.08 \\
UGC02982\_e2 & 04:12:22.15321 & +005.32.48.3227 & 0.21 & 0.31 & 36.15 & 20.72 & 6.77 & 8.00 & 8.90 & 9.50 \\
UGC02982\_e3 & 04:12:22.41490 & +005.32.57.9306 & 0.25 & 0.96 & 36.09 & 20.82 & 6.79 & 8.00 & 8.77 & 5.55 \\
UGC02982\_e4 & 04:12:23.73820 & +005.32.47.1073 & 0.28 & 0.70 & 35.99 & 20.53 & 6.61 & 7.25 & 8.73 & 4.42 \\
UGC03094\_e1 & 04:35:33.74264 & +019.10.24.0963 & 1.45 & 0.84 & 36.79 & 20.74 & 9.00 & 6.34 & 9.49 & 8.16 \\
UGC03094\_e3 & 04:35:33.81702 & +019.09.57.8943 & 1.12 & 0.48 & 36.44 & 20.68 & 9.00 & 6.34 & 9.10 & 3.50 \\
UGC03094\_e2 & 04:35:33.91205 & +019.10.10.0595 & 1.35 & 0.76 & 36.72 & 21.00 & 6.56 & 6.76 & 9.46 & 4.36 \\
IRAS05442+173\_e1 & 05:47:10.82732 & +017.33.46.2608 & 0.91 & 0.31 & 36.16 & 20.95 & 6.56 & 7.02 & 8.76 & 4.08 \\
IC0563\_e2 & 09:46:20.06671 & +003.02.43.6972 & 0.22 & 0.25 & 36.15 & 20.56 & 6.72 & 8.00 & 9.09 & 22.82 \\
NGC3110\_e3 & 10:04:01.52752 & -006.28.26.0344 & 0.62 & 0.78 & 36.39 & 20.87 & 6.61 & 7.22 & 9.15 & 8.45 \\
NGC3110\_e2 & 10:04:02.03336 & -006.28.33.1477 & 1.00 & 0.04 & 36.73 & 21.39 & 6.77 & 8.00 & 9.46 & 6.35 \\
NGC3110\_e1 & 10:04:02.57873 & -006.28.46.3525 & 0.72 & 0.51 & 36.27 & 20.92 & 6.61 & 7.18 & 8.94 & 4.36 \\
NGC3110\_e4 & 10:04:02.68675 & -006.28.35.1127 & 0.40 & 0.83 & 36.22 & 20.66 & 6.61 & 7.18 & 8.99 & 6.73 \\
IC2810\_e1 & 11:25:44.92036 & +014.40.30.8605 & 5.07 & 0.15 & 36.59 & 21.28 & 9.00 & 6.34 & 9.42 & 9.34 \\
NGC5257\_e3 & 13:39:52.19263 & +000.50.22.3494 & 0.68 & 0.37 & 36.12 & 20.97 & 6.62 & 7.39 & 9.03 & 12.09 \\
NGC5257\_e2 & 13:39:52.58999 & +000.50.15.5249 & 1.01 & 0.26 & 36.13 & 21.21 & 6.59 & 8.51 & 9.08 & 3.84 \\
NGC5257\_e1 & 13:39:52.94249 & +000.50.12.6413 & 2.52 & 0.15 & 36.29 & 21.28 & 6.56 & 6.77 & 9.04 & 6.96 \\
NGC5257\_e4 & 13:39:53.54494 & +000.50.28.9336 & 0.56 & 0.66 & 36.28 & 20.92 & 6.58 & 7.49 & 9.11 & 5.34 \\
NGC5258\_e2 & 13:39:57.09557 & +000.49.40.5328 & 1.42 & 0.29 & 36.32 & 21.26 & 6.61 & 7.31 & 9.18 & 6.36 \\
NGC5258\_e3 & 13:39:57.12290 & +000.49.44.0410 & 2.77 & 0.32 & 36.66 & 21.58 & 6.62 & 7.39 & 9.36 & 5.78 \\
NGC5258\_e1 & 13:39:57.23223 & +000.49.47.4126 & 2.20 & 0.52 & 36.77 & 21.62 & 6.74 & 8.00 & 9.52 & 6.18 \\
NGC5331\_e2 & 13:52:16.03444 & +002.06.05.3056 & 4.93 & 0.46 & 37.06 & 21.98 & 6.75 & 8.00 & 9.85 & 2.89 \\
NGC5331\_e4 & 13:52:16.20505 & +002.06.08.5057 & 3.56 & 0.46 & 36.92 & 21.80 & 6.70 & 8.00 & 9.75 & 9.01 \\
NGC5331\_e1 & 13:52:16.30493 & +002.05.59.1366 & 1.06 & 0.81 & 36.64 & 21.24 & 6.65 & 8.85 & 9.44 & 2.70 \\
NGC5331\_e3 & 13:52:16.32488 & +002.06.05.6838 & 5.47 & 0.62 & 37.24 & 22.10 & 6.77 & 8.00 & 10.01 & 4.24 \\
NGC5990\_e2 & 15:46:16.07681 & +002.25.02.0582 & 0.26 & 0.17 & 36.19 & 20.55 & 6.62 & 7.38 & 9.09 & 5.13 \\
NGC5990\_e1 & 15:46:16.52746 & +002.24.47.7330 & 0.63 & 0.60 & 36.29 & 20.81 & 6.56 & 7.06 & 9.13 & 3.55 \\
IRASF16516-09\_e4 & 16:54:23.40443 & -009.53.22.1515 & 0.83 & 0.53 & 36.36 & 21.03 & 6.61 & 7.31 & 9.12 & 4.74 \\
IRASF16516-09\_e1 & 16:54:23.80720 & -009.53.30.6354 & 0.94 & 0.37 & 36.25 & 20.89 & 6.56 & 6.86 & 9.07 & 4.63 \\
IRASF16516-09\_e3 & 16:54:24.67729 & -009.53.15.6416 & 0.31 & 0.69 & 36.04 & 18.65 & 9.00 & 6.34 & 8.77 & 8.90 \\
IRASF17138-10\_e1 & 17:16:35.79572 & -010.20.41.8617 & 7.27 & 0.34 & 37.10 & 22.05 & 6.60 & 8.02 & 9.86 & 4.65 \\
NGC7592\_e1 & 23:18:22.18023 & -004.25.08.0269 & 0.37 & 0.39 & 35.87 & 20.54 & 6.56 & 6.96 & 8.81 & 2.12 \\
NGC7679\_e1 & 23:28:46.50087 & +003.30.43.6612 & 1.48 & 0.68 & 36.71 & 21.37 & 6.59 & 8.96 & 9.38 & 10.38 \\
NGC7679\_e2 & 23:28:46.89743 & +003.30.41.3465 & 0.78 & 0.95 & 36.58 & 20.91 & 6.56 & 7.06 & 9.43 & 0.80
\enddata
\tablenotetext{a}{Star Formation Rates at 33 GHz Calculated using the prescription in Murphy et al. (2012)}
\tablenotetext{b}{Derived $L_{IR}$ and $L_{1.4 GHz}$ given in log SI units (W/Hz)}
\tablenotetext{c}{Starburst99 model ages for each region given in log(age)}
\tablenotetext{d}{Stellar mass given in units of log $M_{\odot}$}
\tablenotetext{e}{Specific Star Formation Rates given in units of $10^{-10}$ yr$^{-1}$}
\end{deluxetable}

\clearpage

\begin{deluxetable}{lllllllllllll}
\tabletypesize{\footnotesize}
\tablecolumns{13}
\tablewidth{0pt}
\tablecaption{Observational Properties of Extranuclear Star-Forming Regions}
\tablehead{
\colhead{Region} & \colhead{$S_{33 GHz}$\tablenotemark{a}} & \colhead{$\sigma_{33 GHz}$} & \colhead{$S_{15 GHz}$} & \colhead{$\sigma_{15 GHz}$} & \colhead{$S_{3 GHz}$} & \colhead{$\sigma_{3 GHz}$} & \colhead{$S_{3.6 \mu m}$\tablenotemark{a}} & \colhead{$\sigma_{3.6 \mu m}$} & \colhead{$S_{4.5 \mu m}$} & \colhead{$\sigma_{4.5 \mu m}$} & \colhead{$S_{8.0 \mu m}$} & \colhead{$\sigma_{8.0 \mu m}$}} \\
\startdata
MCG-02-01-051\_e1 & 0.384 & 0.084 & 0.602 & 0.031 & 1.546 & 0.249 & 0.879 & 0.257 & 0.720 & 0.229 & 12.240 & 2.514 \\
IC1623\_e1 & 0.454 & 0.088 & 0.489 & 0.031 & 1.572 & 0.253 & 0.527 & 0.269 & 0.383 & 0.210 & 5.972 & 2.223 \\
IC1623\_e2 & 0.426 & 0.076 & 0.519 & 0.021 & 1.677 & 0.118 & 1.034 & 0.430 & 0.857 & 0.342 & 16.140 & 4.152 \\
IC1623\_e3 & 0.121 & 0.077 & 0.155 & 0.021 & 0.443 & 0.117 & 0.421 & 0.111 & 0.314 & 0.087 & 6.157 & 1.389 \\
MCG-03-04-014\_e1 & 0.073 & 0.078 & 0.086 & 0.021 & 0.242 & 0.117 & 0.287 & 0.068 & 0.214 & 0.054 & 3.566 & 1.044 \\
NGC0838\_e1 & 0.061 & 0.168 & 0.100 & 0.042 & 0.221 & 0.151 & 0.506 & 0.209 & 0.335 & 0.141 & 5.043 & 0.998 \\
IC0214\_e3 & 0.367 & 0.146 & 0.600 & 0.038 & 1.353 & 0.640 & 0.642 & 0.747 & 0.419 & 0.530 & 13.790 & 3.399 \\
IC0214\_e2 & 0.729 & 0.149 & 0.752 & 0.040 & 2.853 & 0.663 & 0.503 & 0.372 & 0.440 & 0.310 & 28.250 & 2.560 \\
NGC0877\_e2 & 0.492 & 0.146 & 0.183 & 0.039 & 1.882 & 0.659 & 0.370 & 0.250 & 0.371 & 0.291 & 24.520 & 2.220 \\
NGC0877\_e1 & 0.515 & 0.067 & 0.397 & 0.025 & 0.427 & 0.077 & 0.426 & 0.153 & 0.302 & 0.106 & 5.093 & 2.000 \\
UGC02369\_e1 & 0.336 & 0.099 & 0.087 & 0.035 & 0.729 & 0.160 & 0.301 & 0.145 & 0.186 & 0.098 & 6.870 & 3.510 \\
CGCG465-012\_e2 & 0.207 & 0.066 & 0.227 & 0.026 & 0.381 & 0.158 & 0.432 & 0.106 & 0.319 & 0.081 & 5.116 & 0.902 \\
CGCG465-012\_e1 & 0.068 & 0.064 & 0.078 & 0.025 & 0.002 & 0.152 & 0.216 & 0.098 & 0.158 & 0.075 & 3.128 & 1.001 \\
CGCG465-012\_e3 & 0.184 & 0.065 & 0.306 & 0.026 & 0.525 & 0.157 & 0.481 & 0.292 & 0.352 & 0.223 & 6.550 & 2.523 \\
UGC02982\_e5 & 2.622 & 0.086 & 3.248 & 0.038 & 8.807 & 0.236 & 5.084 & 2.175 & 4.530 & 2.011 & 59.130 & 24.14 \\
UGC02982\_e2 & 0.162 & 0.157 & 0.096 & 0.057 & 0.130 & 0.900 & 0.283 & 0.105 & 0.208 & 0.074 & 3.532 & 0.947 \\
UGC02982\_e3 & 0.491 & 0.076 & 0.493 & 0.027 & 2.262 & 0.198 & 1.245 & 1.582 & 0.755 & 1.118 & 30.040 & 9.995 \\
UGC02982\_e4 & 0.372 & 0.110 & 0.628 & 0.041 & 0.877 & 0.286 & 1.465 & 1.486 & 1.141 & 1.131 & 35.150 & 7.380 \\
UGC03094\_e1 & 0.275 & 0.138 & 0.297 & 0.033 & 0.701 & 0.171 & 0.585 & 0.205 & 0.453 & 0.166 & 9.298 & 2.172 \\
UGC03094\_e3 & 0.385 & 0.120 & 0.699 & 0.032 & 2.064 & 0.172 & 1.846 & 0.737 & 1.377 & 0.593 & 26.570 & 4.901 \\
UGC03094\_e2 & 0.237 & 0.125 & 0.316 & 0.033 & 0.627 & 0.174 & 0.907 & 0.168 & 0.673 & 0.139 & 12.270 & 2.050 \\
IRAS05442+173\_e1 & 0.152 & 0.126 & 0.126 & 0.033 & 0.385 & 0.174 & 0.601 & 0.202 & 0.428 & 0.152 & 8.348 & 1.782 \\
IC0563\_e1 & 0.537 & 0.036 & 0.594 & 0.022 & 0.888 & 0.080 & 0.481 & 0.134 & 0.452 & 0.157 & 5.363 & 2.778 \\
NGC3110\_e3 & 0.214 & 0.036 & 0.310 & 0.022 & 0.753 & 0.080 & 0.456 & 0.203 & 0.369 & 0.166 & 3.777 & 1.736 \\
NGC3110\_e2 & 0.146 & 0.036 & 0.157 & 0.022 & 0.442 & 0.080 & 0.359 & 0.219 & 0.247 & 0.168 & 3.628 & 1.542 \\
NGC3110\_e1 & 0.120 & 0.037 & 0.119 & 0.023 & 0.392 & 0.083 & 0.452 & 0.080 & 0.327 & 0.059 & 5.314 & 0.731 \\
NGC3110\_e4 & 0.470 & 0.037 & 0.642 & 0.021 & 1.976 & 0.082 & 1.220 & 0.443 & 0.954 & 0.355 & 16.170 & 4.102 \\
IC2810\_e1 & 0.302 & 0.039 & 0.380 & 0.021 & 0.858 & 0.081 & 0.577 & 0.240 & 0.462 & 0.212 & 5.760 & 2.479 \\
NGC5257\_e3 & 0.590 & 0.039 & 0.794 & 0.022 & 1.784 & 0.082 & 0.899 & 0.405 & 0.748 & 0.327 & 12.690 & 3.619 \\
NGC5257\_e2 & 0.114 & 0.038 & 0.145 & 0.021 & 0.405 & 0.143 & 0.500 & 0.199 & 0.374 & 0.136 & 6.111 & 1.293 \\
NGC5257\_e1 & 0.529 & 0.036 & 0.666 & 0.021 & 2.276 & 0.146 & 1.434 & 0.862 & 1.227 & 0.766 & 16.050 & 6.459 \\
NGC5257\_e4 & 0.586 & 0.036 & 0.847 & 0.021 & 2.977 & 0.149 & 2.030 & 1.178 & 1.697 & 1.033 & 24.070 & 8.083 \\
NGC5258\_e2 & 0.381 & 0.034 & 0.491 & 0.021 & 1.504 & 0.139 & 1.093 & 0.610 & 0.898 & 0.500 & 11.730 & 4.299 \\
NGC5258\_e3 & 0.085 & 0.193 & 0.082 & 0.045 & 0.175 & 0.110 & 0.246 & 0.076 & 0.182 & 0.065 & 2.191 & 0.628 \\
NGC5258\_e1 & 0.669 & 0.105 & 1.062 & 0.031 & 2.308 & 0.093 & 1.805 & 1.110 & 1.334 & 0.853 & 30.120 & 7.059 \\
NGC5331\_e2 & 0.354 & 0.106 & 0.439 & 0.031 & 0.809 & 0.095 & 1.910 & 0.979 & 1.302 & 0.662 & 22.540 & 6.747 \\
NGC5331\_e4 & 0.651 & 0.076 & 1.011 & 0.025 & 3.143 & 0.183 & 1.516 & 1.023 & 1.469 & 0.960 & 17.190 & 9.464 \\
NGC5331\_e1 & 0.097 & 0.096 & 0.233 & 0.025 & 0.519 & 0.172 & 0.595 & 0.296 & 0.439 & 0.224 & 8.395 & 2.875 \\
NGC5331\_e3 & 0.113 & 0.101 & 0.184 & 0.026 & 0.652 & 0.178 & 0.442 & 0.205 & 0.327 & 0.149 & 7.221 & 2.184 \\
NGC5990\_e2 & 0.125 & 0.115 & 0.134 & 0.027 & 0.340 & 0.178 & 0.398 & 0.058 & 0.287 & 0.049 & 5.819 & 0.789 \\
NGC5990\_e1 & 0.090 & 0.101 & 0.139 & 0.026 & 0.402 & 0.175 & 0.413 & 0.183 & 0.299 & 0.133 & 4.952 & 1.444 \\
IRASF16516-09\_e4 & 0.286 & 0.093 & 0.322 & 0.025 & 0.236 & 0.206 & 1.010 & 0.361 & 0.749 & 0.262 & 15.970 & 4.950 \\
IRASF16516-09\_e1 & 0.265 & 0.096 & 0.311 & 0.026 & 0.434 & 0.211 & 0.962 & 0.206 & 0.718 & 0.174 & 13.360 & 2.582 \\
IRASF16516-09\_e3 & 0.220 & 0.110 & 0.324 & 0.026 & 0.209 & 0.208 & 0.447 & 0.146 & 0.364 & 0.121 & 6.986 & 1.627 \\
IRASF17138-10\_e1 & 0.422 & 0.068 & 0.413 & 0.024 & 0.963 & 0.102 & 1.488 & 0.349 & 1.080 & 0.276 & 16.980 & 3.557 \\
NGC7592\_e1 & 0.175 & 0.068 & 0.215 & 0.024 & 0.528 & 0.102 & 1.339 & 0.221 & 0.955 & 0.168 & 13.660 & 2.446 \\
NGC7679\_e1 & 0.236 & 0.061 & 0.183 & 0.024 & 0.322 & 0.092 & 0.410 & 0.093 & 0.332 & 0.093 & 5.897 & 1.529 \\
NGC7679\_e2 & 0.029 & 0.063 & 0.142 & 0.025 & 0.272 & 0.094 & 0.621 & 0.135 & 0.471 & 0.131 & 8.380 & 1.737
\enddata
\tablenotetext{a}{All VLA and Spitzer IRAC fluxes given in mJy}
\end{deluxetable}

\end{document}